%% file: confusing_v2.tex
\documentclass[aps,prd,amsmath,floats,floatfix,twocolumn,superscriptaddress,nofootinbib,noshowpacs]{revtex4}

\usepackage{amssymb}
\usepackage{amsmath}
\usepackage{verbatim}
\usepackage{mathrsfs}
\usepackage{amsfonts}
\usepackage{latexsym}
\usepackage{epsfig}
\usepackage{color}
\usepackage{graphicx,subfigure}
\usepackage{units}
\usepackage{overpic}
\usepackage[utf8]{inputenc}
\usepackage{ulem,xpatch}

\usepackage{enumerate}
\usepackage{pgfplots}

\begin{document}
\title{Confusing dark matter particle properties with modifications to General Relativity}
\author{Armando A. Roque}
\author{J. Barranco}
\affiliation{Division de Ciencias e Ingenier\'ias,  Universidad de Guanajuato, Campus Leon, C.P. 37150, Le\'on, Guanajuato, M\'exico.}
\date{\today}
\begin{abstract}
Cold Dark Stars made of self-gravitating fermions in the degenerate limit are constructed in General Relativity and in R-square gravity,
$f(R)=R+\alpha R^2$. The properties of the resulting Cold Dark Stars in both theories of gravity are studied. It is found that the same gravitational potential is generated for different election of the parameters of the model, such as the mass of the fermion, the self-interacting strength or the value of $\alpha$, thus, a possible confusion in the determination of the dark matter properties and the favored theory of gravity might arise.
\end{abstract}

\maketitle

\section{Introduction}
Current astrophysical observations favor General Relativity (GR) as the correct theory that describes the gravitational interaction~\cite{Abbott:2018lct,Sakstein:2017xjx}. Nevertheless, if GR is valid at cosmological scales, in order to have a concordance model for the evolution of the universe two dark components must be added to the energy density content of the universe i.e. Dark Matter (DM) and Dark Energy (DE). The latest in the form of a cosmological constant. The model of the evolution of the universe that have as main ingredients GR, DM as a cold heavy non relativistic particle and a cosmological constant is known as the $\Lambda$-CDM model~\cite{Peebles1993PrinciplesOP}. The predictions of $\Lambda$-CDM model are consistent with most of the  observational data such as the measurement of anisotropies in the temperature and polarization of the cosmic microwave background (CMB)~\cite{Bennett:2013}, fluctuations in the density of baryonic matter (BAO)~\cite{Eisenstein:2005su, Cole:2005sx, Alam:2016hwk}, observations of the magnitude-redshift relation for high redshift SNe Ia~\cite{Riess:1998cb, Scolnic:2017caz}) among many other observations. However, there are some unresolved problems where the predictions are in tension with the observations. For instance, the Planck best-fit measurements of the current expansion rate $H_0$~\cite{Akrami:2018vks} is in tension with  the value obtained by local measurements of $H_0$~\cite{Riess:2019cxk, Wong:2019kwg, Freedman:2019jwv, Schoneberg:2019wmt}.  Furthermore, the $\Lambda$-CDM model is not  compatible with some astrophysical observations at galactic scales. Perhaps more important, $\Lambda$-CDM model has little to say about the nature of dark matter, except that it could be a heavy neutral particle that interacts very weakly with the rest of the particles of the Standard Model of particle physics (SM).

In order to reduce those problems, an interesting possibility emerges: that besides adding dark matter and dark energy as main components of the energy density of the universe, perhaps GR needs to be modified or extended in such a way that there are not inconsistencies at cosmological and galactic scales. In this scenario, an unpleasant possibility is that {\it there could be a degeneracy in the determination of DM properties once variations to the theory of General Relativity are allowed}. The objective of this work is to show one example of this possibility. 

We will work out this example in a simplified model where DM particles interact only through gravitational interactions, and thus, the only parameters to be determined are the spin and the mass of the dark matter particle, plus the coupling constants that parametrize the interactions among themselves. If DM has no other interaction than gravitational, then DM particle properties will be obtained only through astrophysical observations of the dynamics of visible objects with the gravitational force generated by the DM.

For definitiveness, we will consider as a model for DM particles, fermions of mass $m$, without interactions with the standard model of particles except the gravitational interaction. In particular this model of fermionic dark matter has been used to model dark matter halos~\cite{Destri:2012yn,Destri:2013pt,Domcke:2014kla,Randall:2016bqw,Barranco:2018gjg,Savchenko:2019qnn,Gomez:2019mtl}. This dark matter fermions can interact themselves and thus a self-interaction term could be added~\cite{PhysRevD.64.043005}. 

On the other hand, as for modification of GR, we will focus on a theory of gravity that is derived from the action
\begin{equation}
S=- \frac{1}{16\pi G}\,\int d^{4}x \sqrt{-g}\,(\text{R}+\alpha \text{R}^{2})\,, \label{action}
\end{equation}
where $\text{R}$ is the Ricci scalar and the $\alpha$ is a constant with units of the inverse of the Ricci scalar~\footnote{We will use unit where $\hbar=c=1$.}. This model is known as R-square gravity or as the Starobinsky model~\cite{Starobinsky:1980te,Gottlober:1990um,Cembranos:2008gj}. This modification to GR is a particular case of the so-called $f(R)$ theories of gravity~\cite{Capozziello:2011et}. Note that R-square gravity is not introduced to solve the low energy problems of GR such as to avoid the introduction of dark matter or dark energy. Instead, we have used it because R-square gravity  it is the simplest non trivial four-derivative extension of GR that is free of ghosts~\cite{Cembranos:2008gj,Capozziello:2011et} and thus it is a natural extension to GR to be studied. We have been motivated to use this fermionic dark matter model and R-square gravity because the minimal number of parameters that are introduced. Even in this minimal scenario we will show there could be a confusion in the determination of fermions properties for different values of $\alpha$. In general, other modifications to GR (see for example~\cite{PhysRevD.76.064004, Starobinsky:2007hu, Linder:2009jz, Odintsov:2017qif}) have more free parameters, thus strengthening the possible confusion in the determination of DM particle properties.

For a distant observed, the gravitational effect of compact objects on test particles depend only of their total mass and radius (i.e. the compactness) of the compact self-gravitating configuration. What we will show is that self-gravitating configurations made of DM fermions will have the same mass and radius although in one case one configuration is made of fermions with a specific value of self-interacting coupling constant in General Relativity while the equivalent configuration is made for a different self-interacting constant but in R-square gravity. Thus, there is a possible confusion in the determination of the dark matter properties. Even in this very simplified scenario where few parameters are present. In the case where DM-SM interactions are introduced, a bigger indetermination will be
expected.  

The paper is organized as follows: In Section~\ref{configurations} we obtain the self-gravitating objects made of fermionic dark matter,  both in General Relativity and in R-square gravity. Some previous results in GR are reproduced~\cite{Narain:2006kx} and new results are obtained, specially those concerning the compactness of the resulting configurations. We present in section~\ref{sec:results} a comparison of those configurations in GR and \mbox{R-square} gravity and the equivalence of some configurations even for different values of the self-interacting coupling, i.e. a confusion on the determination of the DM properties. In Section~\ref{sec:conclusions} we give some concluding remarks.

\section{Cold Dark Stars}\label{configurations}
\begin{figure}
	\centering
	\scalebox{0.41}{
		\input{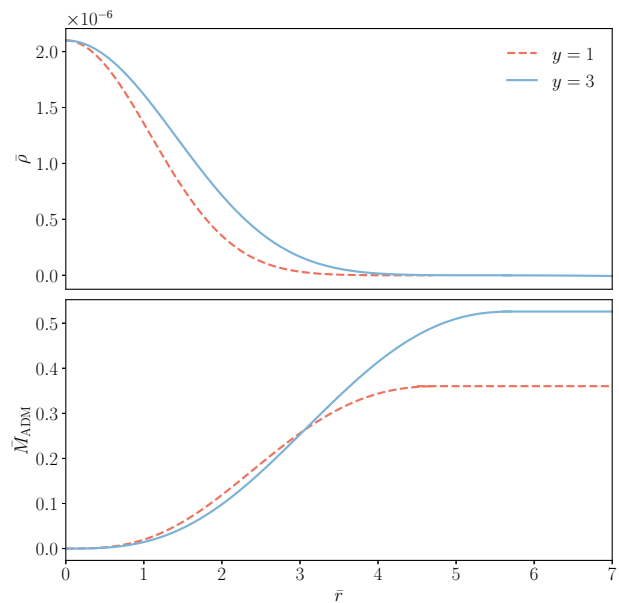}
	}
\caption{Typical energy density profile and mass as a function of the radial coordinate $r$. For definitiveness it was chosen $\bar{\rho}_{0}=2.1\times10^{-6}$, and two values of the self-interacting strength: $y=1$ (dashed line) and $y=3$ (solid line). The introduction of self-interaction produces a more massive self-gravitating structure and with a stiffer density profile.}\label{Fig01}
\end{figure}

\begin{figure}
	\centering
	\scalebox{0.33}{\input{RG.pgf}} 
	\caption{ Black solid line correspond to $\bar{M}_{\mathrm{ADM}}$ {\it vs} $\bar{R}$  for $y=0$ (General Relativity without auto-interaction) and the blues doted-lines corresponds to the dark star made of interacting fermions in GR with coupling constant for a rang between $y=1$ and $y=5$.}\label{Fig0}
\end{figure}
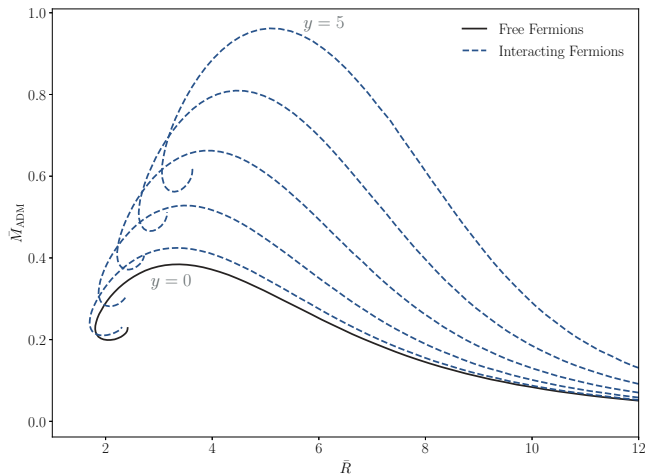
Dark matter over-densities might form small clumps that can evolve into Dark Stars, i.e self-gravitating objects made of dark matter~\cite{Freese:2008hb,Berezinsky:2014wya,Freese:2015mta}. In is generally assumed that those Dark Stars will be powered by the heat from dark matter annihilation, rather than by fusion. In our case, since we are assuming that DM has no interaction with SM particles, thus, our Dark Stars will be Cold compact objects and in order to distinguish from dark stars we will call the resulting self-gravitating objects made of Fermionic DM in the degenerate limit as Cold Dark Stars (CDS). 

\subsection{CDS in General Relativity}
For non interacting fermions in the degenerate limit, it is possible to establish a relationship between the pressure and the density, for a gas of free fermions. This relationship can be calculated via explicit expressions for the energy density $\rho$ and pressure $p$. For a completely degenerate gas of fermions $\rho$ and $p$ are given by~\cite{Landau:1980mil, Narain:2006kx}
\begin{equation}
\rho(z)=\frac{m^4}{8\pi^{2}}\left[ \left(2 z^{3}-3z\right)\left(1+z^{2}\right)^{1/2}+3\sinh^{-1}(z)\right]\,,
\end{equation}

\begin{equation}
p(z)=\frac{m^4}{24\pi^{2}}\left[ \left(2 z^{3}+z\right)\left(1+z^{2}\right)^{1/2}-\sinh^{-1}(z)\right]\,,
\end{equation}
where $z=k_{f}/m$ is the dimensionless Fermi momentum and $m$ the mass of the fermion. 
It is convenient to work in dimensionless variables so we define the dimensionless variables
\begin{equation}
	\bar p=\frac{p}{m^4},\;\; \bar \rho=\frac{\rho}{m^4},\;\;\bar{r}=r \frac{m^2}{m_p}\,, \label{dimensionless}
\end{equation}
where, $m$ is the fermion mass, and $m_{p}$ is the Planck mass, defined as $m_{p}=\text{G}^{-1/2}$. 

We will consider self-interacting dark matter since it may resolve some  problems of $\Lambda$-CDM paradigm at galactic scales~\cite{Spergel:1999mh}. One of the simplest Lagrangian for this kind of fermionic interaction is
 \begin{equation}
 \mathcal{L}=-g \bar\chi \chi H\,,\label{lagrangian}
 \end{equation} 
where $H$ is the particle that mediates the interaction between them, with mass $m_H$, and  $g$ is a coupling constant. The fermionic dark matter particles $\chi$ has mass $m$. Following~\cite{Narain:2006kx}, in a mean field approximation, the interparticle interactions given for such interaction can be added effectively by including a term proportional to the square of the density number of fermions as follows
\begin{eqnarray}
	\bar{\rho}(z)&=&\frac{1}{8\pi^{2}}\left[ \left(2 z^{3}+z\right)\left(1+z^{2}\right)^{1/2}-\sinh^{-1}(z)\right]\nonumber\\ & &\;\;\;+\frac{1}{9\pi^{4}}y^{2}z^{6}\,, \label{eqrho}
\end{eqnarray}
\begin{eqnarray}
	\bar{p}(z)
	&=&\frac{1}{24\pi^{2}}\left[ \left(2 z^{3}-3z\right)\left(1+z^{2}\right)^{1/2}+3\sinh^{-1}(z)\right]\nonumber\\ & &\;\;\;+\frac{1}{9\pi^{4}}y^{2}z^{6}\,,\label{eqp}
\end{eqnarray}
where  we have defined $y=g\, m/m_{H}$ as the effective dimensionless constant that parametrize the interaction strength between fermions in terms of the coupling $g$, the mass of the mediator $m_{H}$, and the mass of the dark matter fermion $m$. 
Thus, this model for dark matter has only two free parameters $m$ and $y$. Note that given the Lagrangian~\eqref{lagrangian}, it is possible to estimate the possible values of $y$ allowed from astrophysical observations. 

The  self-interacting cross-section for dark matter fermions with a interaction given by the Lagrangian~\eqref{lagrangian} is 
\begin{equation}
\sigma=\frac{g^4}{8\pi m_H^4}m^2=\frac{1}{8\pi}\frac{y^4}{m^2}\,.\label{cs}
\end{equation}
The tightest observational constraints on dark matters interaction cross-section come from the lack of deceleration of dark matter in the cluster collisions. Statistical analysis of 72 of such collisions puts a restrictive bound $\sigma/m <0.47 \mbox{g/cm}^2$~\cite{Harvey:2015hha}.
This bound and the cross section given by eq.~\eqref{cs} gives an upper bound on the interaction strength $y$ as a function of the mass of the dark matter fermion $m$
\begin{equation}
y<15.24\left(\frac{m}{\mbox{GeV}}\right)^{3/4}\,.\label{bound}
\end{equation}
Thus, for $m=1$ TeV, the dimensionless variable $y$ can be as big as $\sim 1500$. 

Next, in order to obtain the self-gravitating object made of this gas of fermions, we solve the Einstein's equations and the conservation of the energy tensor. For $\alpha=0$, the action eq.~\eqref{action} reduces to the General Relativity action. We consider a static, spherically symmetric ansatz for the line element
\begin{equation}
ds^{2}=-A^{2}(r)dt^{2}+B^{2}(r)dr^{2}+r^{2}d\theta^{2}+r^{2}\sin^2\theta d\varphi^{2}\,\label{metric}.
\end{equation}
The energy momentum tensor for a perfect fluid is given by
\begin{equation}
T^{\mu\nu}=(\rho+p)u^{\mu}u^{\nu}+g^{\mu\nu}p\,,
\end{equation}
and thus, Einstein-equations reduces to the Tolman-Oppenheimer-Volkoff equations (TOV):
\begin{eqnarray}
	B'&=&\frac{B^3 \left(8\pi\bar{r}^2 \bar{\rho}-1\right)+B}{2 \bar{r}},\\
	2 A'&=&\frac{A \left(B^2 \left(8\pi\bar{r}^2 \bar{p}+1\right)-1\right)}{\bar{r}},\\
	\bar{p}'&=&-\frac{A' (\bar{\rho}+\bar{p})}{A}\,.
\end{eqnarray}
We set as boundary conditions at $r=0$
\begin{eqnarray}
A(0)&=&B(0)=1\,,\nonumber\\
\bar{p}(0)&=&\bar p_{0}\,,\label{bc}
\end{eqnarray}
with $\bar p_{0}$ a free parameter. 
As the equation of state depends on two parameters, $z$ and $y$, the $\bar{\rho}(\bar{p})$ relationship was obtained by interpolating eqs.~\eqref{eqrho} and eq.~\eqref{eqp} for a range of values $0<z<20$ given a fixed interaction strength $y$. The range of \mbox{values} explored for $y$ are between $0$ and $10$, consistent with the upper bound expressed by eq.~\eqref{bound} for $m\sim 1$ GeV. 
In our analysis, the central pressure of the object $\bar{p}_{0}$  took values between $10^{-10}$ to $10$.

The mass of this objects was calculated using the ADM mass
\begin{eqnarray}
M_{\text{ADM}} &=&\frac{r}{2 G}\left(1-\frac{1}{B^2}\right)\,,\nonumber\\
&=&\bar{M}_{\text{ADM}}\frac{m_p^3}{m^2}\,.
\label{eqmas}
\end{eqnarray}

It is convenient to define the radius $\bar{R}$ as the point where $\bar p(\bar r) <0$, and this point will represent the radius of  CDS. 
In this way, we fix $\bar p(\bar r > \bar R)=0$ and then the ADM mass will be constant for $\bar r >\bar R$. Thus, each CDS will have a fixed total mass $\bar{M}$ and a finite radius $\bar{R}$. Typical configurations are shown in Fig.~\ref{Fig01}, where the profile for the energy density as a function of $\bar{r}$ are shown for a fixed value of $\bar{p}_0$ (that is equivalent to a fixed value of $\bar{\rho}_0$) for two different values of the self-interacting strength $y$. For definitiveness it was chosen $\bar{\rho}_0=2.1\times10^{-6}$, and two values of the self-interacting strength, namely $y=1$ (dashed line) and $y=3$ (solid line). In general, CDS with self-interacting fermions are more massive  and with a stiffer density profile than those self-gravitating structures with $y=0$.  

It is possible to obtain the full set of CDS for our parameter space $(\bar p_0,y) \to (\bar{R}, \bar{M}, y)$. These configurations are shown in Fig.~\ref{Fig0}. 
Note that there are different branches of configurations because each value of $y$ implies a different equation of state. For small values of $y$ and low densities, the interaction terms can be ignored and  the object can be described for an ideal degenerate Fermi gas. Because the upper bound we found in eq.~\eqref{bound}, only small values of $y$ are allowed for small values of $m$. 

For higher $y$ values, only accessible for high values of the fermion mass, the interaction terms become more and more important. 
The maximum mass grows for higher values of $y$. The growth in the mass is produced by the increase in the pressure due to the inclusion of the self-interacting term and thus the degeneration pressure is strengthened. This increase in the pressure requires to increase the total mass of the self-gravitating object such as the gravity counteracts the pressure in order to achieve hydrostatic equilibrium.

Figures~\ref{Fig01}-\ref{Fig0} are presented in the dimensionless variables eq.~\eqref{dimensionless}. In order to restore physical quantities, the following relations are needed
\begin{eqnarray}
M_{phys}&=&1.638 \times 10^{12}\, \bar{M}_{\text{ADM}}\left(\frac{m}{\text{KeV}}\right)^{-2} \;M_{\odot}\,,\label{physical}\\
R_{phys}&=&0.078\,\bar{R}\left(\frac{m}{\text{KeV}}\right)^{-2}\;\text{pc}\,,\\
\rho_{phys}&=&2.3\times 10^{-4}\;\bar{\rho}\; \left(\frac{m}{\text{KeV}}\right)^4\;\frac{\text{kg}}{\text{m}^{3}}\,.\label{physica3}
\end{eqnarray}

A full set of different astrophysical objects with different masses and radius are possible. For example, if $m=1~$KeV then $M_{phys}\sim10^{12}M_{\odot}$ and $R_{phys}\sim 0.1$ pc, e.g. a possible super massive black hole mimicker. Although this possible black hole mimicker will not have event horizon and thus they are more similar to Boson Stars~\cite{Torres:2000dw,Guzman:2009zz,Liebling:2012fv} or Gravastars~\cite{Mazur:2004fk}. On the other hand, if $m \sim 1$ TeV, then $M_{phys}\sim10^{-6}M_{\odot}$ and $R_{phys}\sim 1$ cm.
Then, CDS will have properties similar to axion stars~\cite{Barranco:2010ib, Eby:2014fya, Braaten:2019knj}.

We finish this section by enumerating some general properties of CDS.
In GR the self gravitating configurations made of degenerate non interacting fermions have the following properties
\begin{enumerate}
\item The total mass of the CDS increases as $\bar \rho_0$ increases. There is a critical value of 
$\bar \rho_\star$ where CDS reaches a maximum value of the mass, $\bar{M}_{max}$. For $\rho_0>\rho_\star$ the total mass  $M$ is smaller that $\bar{M}_{max}$.  
\item The introduction of a self-interacting coupling between the fermions ($y>0$) make that the resulting CDS increases the value of the maximum mass $\bar{M}_{max}$.
\item The radius of the configuration $\bar{R}$ decreases as $\bar \rho_0$ increases.
\end{enumerate} 
Next we will study CDS in R-square gravity in order to find how their properties change.  

\subsection{Cold Dark Star in R square gravity}
\begin{figure*}
	\centering
	\scalebox{0.45}{\input{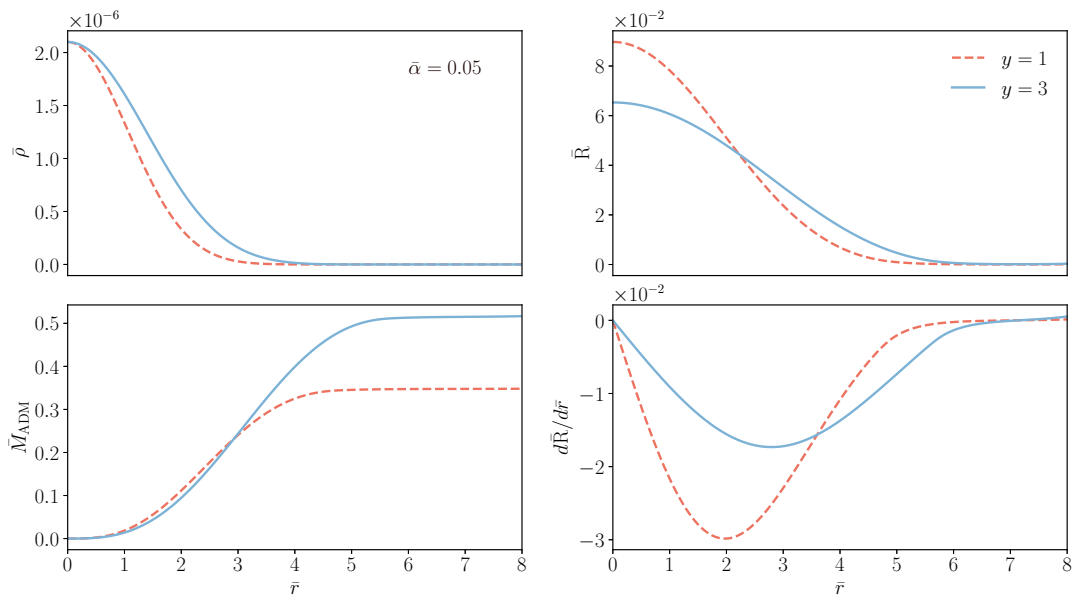}}
	\caption{Left panel: CDS density and Mass profile in R-square gravity with $\bar{\alpha} = 0.05$. 
		As in Fig. 1, $\bar \rho_{0} =2.1\times 10^{-6}$ and 
		we have selected the values of the interaction strength $y=1$ and $y=3$. Right panel: The value of Ricci scalar in the origin was chosen so that asymptotically tend to zero.}\label{Fig03}
\end{figure*}
Now we study CDS in the R-square gravity given by the action eq.~\eqref{action} ($\alpha \ne 0)$. 
The general field equations (the equivalent to the Einstein equations) for a $f(R)$ theory of gravity are given by
\begin{eqnarray}
	f_{R} R_{\mu\nu}-\frac{1}{2} g_{\mu\nu} f(R)+ g_{\mu\nu}\Box f_{R} -
	\nabla_{\nu}\nabla_{\mu} f_{R}= k T_{\mu\nu},\label{emod}
\end{eqnarray}
where, $f_{R}:=\partial_{R} f(R)$, $\Box=g^{\mu\nu}\nabla_{\mu}\nabla_{\nu}$.
As in the previous case, we will work in a spherically symmetric line element eq.~\eqref{metric}, with $T^{\mu\nu}$ a perfect fluid with equation of state given by eqs.~\eqref{eqrho} and \eqref{eqp}. It is convenient to define the new dimensionless variables $\bar{\alpha}$, $\bar{R}$ as follows
\begin{eqnarray}
\alpha=\bar{\alpha}\,m_{p}^2/m^4\,,\;\;\;
\text{R}=\bar{\text{R}}\, m^{4}/m_{p}^{2}\,.
\end{eqnarray}
Taking $f(\text{R})=\text{R}+\alpha \text{R}^2$, in eq.~\eqref{emod} and using the 1-1, \mbox{2-2} components and the conservation equation 
\mbox{$\nabla_{\mu} T^{\mu \nu}=0$} with as before, it is possible to obtain the modified Tolman-Oppenheimer-Volkoff for R-square gravity. 
The R-square TOV system is given by
\begin{widetext}
	\begin{eqnarray}
	B'&=&\frac{B \left(2+B^2 \left(-2+\bar{\alpha} \bar{\text{R}} \left(-4+\bar{r}^2 \bar{\text{R}}\right)+16 \pi \bar{r}^2 \bar{\rho}\right)+4 \bar{\alpha}\left(\bar{\text{R}}+\bar{r}\left(2 \bar{\text{R}}'+\bar{r} \bar{\text{R}}''\right)\right)\right)}{4 \bar{r} \left(1+2 \bar{\alpha} \bar{\text{R}}+\bar{r} \bar{\alpha}\bar{\text{R}}'\right)}\,,\label{eq0}\\
	A'&=&\frac{A \left(-2-4 \bar{\alpha}\bar{\text{R}}+B^2 \left(2+16\pi\bar{r}^2\bar{p}+\bar{\alpha}\bar{\text{R}} \left(4-\bar{r}^2\bar{\text{R}}\right)\right)-8\bar{r}\bar{\alpha} \bar{\text{R}}'\right)}{4 \bar{r} \left(1+2 \bar{\alpha} \bar{\text{R}}+\bar{r} \bar{\alpha}\bar{\text{R}}'\right)}\label{eq01}\,,\\	
	\bar{p}'&=&-\frac{A' (\bar{\rho}+\bar{p})}{A}\,.\label{eq02}
	\end{eqnarray}
\end{widetext}
\hfill \break

In this case, an extra equation that describe the Ricci scalar behaviour arises because the theory has an extra degree of freedom. It is given by
\begin{widetext}
\begin{equation}
	\bar{\text{R}}''=\frac{6\bar{\alpha}\bar{\text{R}}'\left(-1-2\bar{\alpha} \bar{\text{R}}+2\bar{r}\bar{\alpha}\bar{\text{R}}'\right)+B^2 \left(\bar{r}(1+2\bar{\alpha}\bar{\text{R}})(24\pi  \bar{p}+\bar{\text{R}}-8\pi\bar{\rho})+\bar{\alpha}\left(-6+\bar{\text{R}} \left(\bar{r}^2-12\bar{\alpha}+3\bar{r}^2 \bar{\alpha} \bar{\text{R}}\right)+16 \pi  \bar{r}^2 \bar{\rho}\right) \bar{\text{R}}'\right)}{6\bar{r}\bar{\alpha}(1+2\bar{\alpha} \bar{\text{R}})}\,.\label{eq10}
\end{equation}
\end{widetext}

The value of the free parameter $\alpha$ of R-squared theory is constrained from observations in different scales. In the cosmological context, if one takes $\alpha < 0$ then ghost modes instabilities arises~\cite{Barrow_1983}. Furthermore, for negative values of $\alpha$, the Ricci scalar profile has a oscillating behaviour outside the star. Similar behaviour occurs in the neutron star context~\cite{Astashenok_2017}. Then, we will work with $\alpha > 0$ values.

In the strong gravity regime, $\left| \alpha \right| $ is constrained to be  $\lesssim 10^{10} \;\text{cm}^{2}$~\cite{Arapo_lu_2011}. For weak-field limit, it is constrained by different experiments: E$\ddot{\text{o}}$t-Wash experiments provides the more stringent bound to be $\left| \alpha \right| \lesssim  10^{-6}\;\text{cm}^{2}$. Furthermore, the Gravity Probe B experiment constrains $\left| \alpha \right|$ to have values $\left| \alpha \right| \lesssim  5\times 10^{15}\;\text{cm}^{2}$ and from measurements of the precession of the pulsar B PSR J0737-3039 constrains $\left| \alpha \right| \lesssim 2.3\times10^{19} \; \text{cm}^{2}$~\cite{PhysRevD.81.104003}. Although all the bounds are different, all they are still meaningful, because this type of theory present a chameleon effect and thus the $\alpha$ values could be different at different scales. 

In our case, to compare $\alpha$ with $\bar \alpha$ we recall that,
\begin{equation}
\alpha= 5.79 \times 10^{-5} \bar \alpha \left(\frac{m}{1\mbox{TeV}}\right)^{-2} \mbox{cm}^2\,.
\end{equation}
For $m \sim 1$ TeV, values of $\bar \alpha \sim 10^{-1}$ will be in concordance with the strongest limit of $\left| \alpha \right| \lesssim  10^{-6}\;\text{cm}^{2}$. We will find the  configurations for a specific value of $\bar \alpha=0.05$ in order to be consistent will all current bounds.

The next step for the numerical analysis is to define the boundary conditions compatible with solutions that are regular at the origin. In addition to the boundary conditions given by eqs.~\eqref{bc} for the metric components $A$ and $B$, and $\bar p$ at $\bar r=0$, it is needed to establish boundary conditions for  the Ricci scalar $\bar{\text{R}}$ and its derivative. By doing an expansion the system equation around to the origin it was found as boundary conditions 
\begin{align}\label{boundayC}
\bar{\text{R}}(0)=\text{R}_{0},\;\;\; \bar{\text{R}}'(0)=0\,,
\end{align}
where the ``prime'' indicate derivation with respect to $\bar{r}$. Here $\text{R}_{0}$ has to be chosen so that the Ricci scalar vanishes asymptotically (at $r \to \infty$). The correct value of $\text{R}_{0}$ such as the boundary conditions are fulfilled is found with a shooting algorithm~\cite{Numerical}. Furthermore, we will impose that the solutions must be localized and asymptotically flat. The Ricci scalar decays as $\bar r^{-2}$ for for $\bar r\to \infty$, that is $\text{R}$ approaches asymptotically to zero. In order to achieve this   numerically, we have to choose a particular value of $r$ such that the value of Ricci scalar is almost zero. Numerically we had chosen a value $\bar{r}=r_{\star}$ such as \mbox{$\bar{\text{R}}(r_{\star})\le \text{R}_{0}/10^{4}$}.

The CDS mass in R-square gravity is computed with eq.~\eqref{eqmas} evaluated at $\bar{r}=r_{\star}$ because the Ricci scalar is negligible and then the space-time metric approaches to the Schwarzschild space time. Another important observation is that the density tends to zero faster that the Ricci scalar, and then, the ADM-mass can increases even if the star has negligible contribution of fermions. The same holds for the derivative of the Ricci scalar. Thus, the fermions are contained in a radius that is smaller than $\bar r_\star$. 

The system of equations~\eqref{eq0}-\eqref{eq10} were solved numerically with an equation of state given by eqs.~\eqref{eqrho}-\eqref{eqp}, i.e.the equation of state for a gas of self-interacting fermions in the degenerate limit. Typical CDS configurations in R-square gravity for $\bar \alpha=0.05$ are shown in Fig.~\ref{Fig03}. As an example, two different values of the coupling strength ($y=1$ red solid line and $y=3$ blue line) are plotted. It is shown the density, $\bar{\rho}(\bar{r})$, and mass, $\bar{M}(\bar{r})$, profiles for the specific value of the central pressure $\bar \rho_0=2.1\times10^{-6}$. For completeness, the profile for the Ricci scalar and their derivative are also shown in the right panel of Fig.~\ref{Fig03}. 
\begin{figure}
	\centering
	\scalebox{0.33}{\input{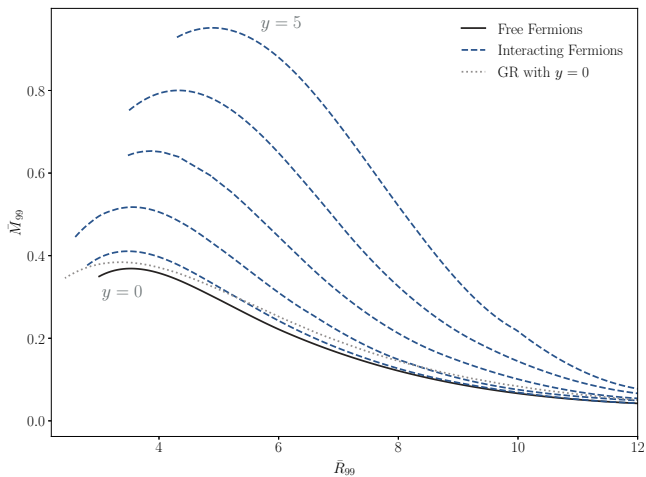}}
		\caption{As in Fig.~\ref{Fig0}, we obtained CDS configurations, but now in a R-square gravity with $\bar{\alpha} = 0.05$. Black solid line correspond to $y=0$ and the blues doted-lines corresponds to the dark star made of self-interacting fermions with coupling constant between $y=1$ and $y=5$. Grey line correspond to CDS in GR with $y=0$ in order to compare GR and R-square CDS configurations.}\label{Fig05}
\end{figure}

We finish this section by constructing all configurations for a specific value of $\bar \alpha=0.05$. Solutions in the $M$ {\it vs} $R$ space for different values of the self-interacting strength $y$ are shown in Fig.~\ref{Fig05}.

In R-square gravity the CDS have the same behaviour as in GR, but the new parameter $\bar{\alpha}$ modifies the global structure of the configuration. In particular, in R-square gravity the maximum masses $M_{max}$ are smaller than the GR case for the same value of the self-interacting coupling constant $y$ (see Figs.~\ref{Fig0}-\ref{Fig05}). In order to understand this behaviour, let us take the Newtonian limit of \mbox{R-square} gravity.

In this case, the Newtonian gravitational potential for R-square gravity is~\cite{Stelle:1977ry}:
\begin{equation}
V(r)=-\frac{GM}{r}-\frac{GM}{3r}e^{-\beta r}\,,\label{newtonian}
\end{equation} 
where $\beta=\frac{1}{2}\sqrt{3 \alpha}$. Thus, the gravitational force is increased if $\beta>0$, and consequently, a lower amount of mass is needed to compensate the pressure produced by the fermions. This reduction in the mass of the configuration can be seen in Fig.~\ref{Fig05} where we have included in a grey solid line the corresponding CDS configurations for $y=0$ in GR. The black solid line that corresponds to R-square gravity is always below the grey line, meaning that all CDS in R-square gravity are smaller that the GR case.   

\section{Similarities and differences of CDS in GR and R-square gravity} \label{sec:results}

\subsection{Compactness}
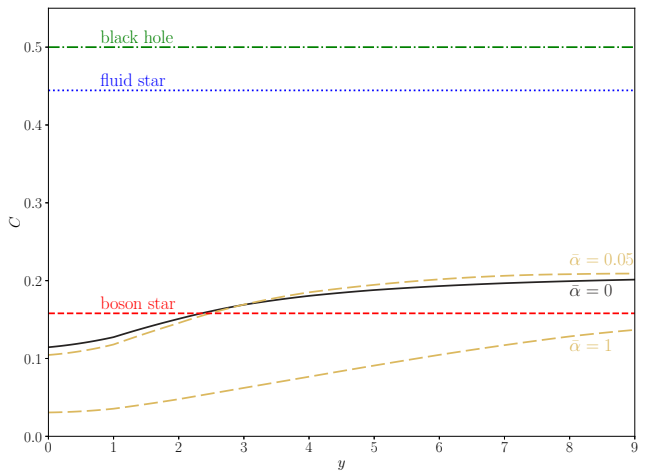
\begin{figure}
	\centering
	\scalebox{0.33}{\input{Compac_RG.pgf}}
		\caption{Compactness of Fermionic dark stars as a function of the self-interaction strength parameter, y. The black solid line correspond to the case $\bar{\alpha}=0$ (General Relativity), the orange dashes lines represent the compactness for two case, $\bar{\alpha}=0.05$, and $\bar{\alpha}=1$.}\label{compactness}
	\end{figure}

The Cold Dark Stars are objects make of self-interacting fermionic dark matter that do not interact with SM particles. For this reason they can not be observed by standard electromagnetic probes. However, in principle, they can be observed by gravity probes, such as gravitational wave (GW) signals or by the effect that CDS can induce gravitationally to other stars (e.g.,~\cite{Palenzuela:2017kcg, Bezares:2018qwa}). 
GW signals of horizonless objects, as it is the case of CDS, will have readily distinguishable behaviour as compared with GW signals of those compact objects that indeed have an event horizon \cite{Kesden:2004qx}. In other words, GW detection will be the ultimate confirmation of the true nature of compact objects. 

Nevertheless, in the later case, an object that is far to the Cold Dark Star will feel the same gravitational potential both in GR and in R square gravity as long as the ratio $M/R$ is the same. The parameter $M/R$ is called the compactness of the star $C$ defined as
\begin{equation}
C=\frac{\bar{M}_{ADM}}{\bar{R}}=\frac{GM_{phys}}{R_{phys}c^2}\,,
\end{equation}
where $G$ is the gravitational constant, $c$ the speed of light, and $M_{phys}$, $R_{phys}$ are defined in eq.~\eqref{physical}-\eqref{physica3}. Note that $C$ is a dimensionless parameter. 

The compactness of CDS is affected when we considered a self-interacting fermionic dark matter: the larger the value of $y$ the bigger the mass, meanwhile the radius of the configuration is not severely affected by the value of $y$ and thus, the compactness of the star increases as $y$ increases. Changes in the compactness will imply changes in several properties such as the possible gravitational radiation emitted by an asymmetric star, or a binary of CDS and of course a change in the compactness induces changes in the gravitational potential the stars produces in other objects.
\begin{figure}
	\centering
	\scalebox{0.33}{\input{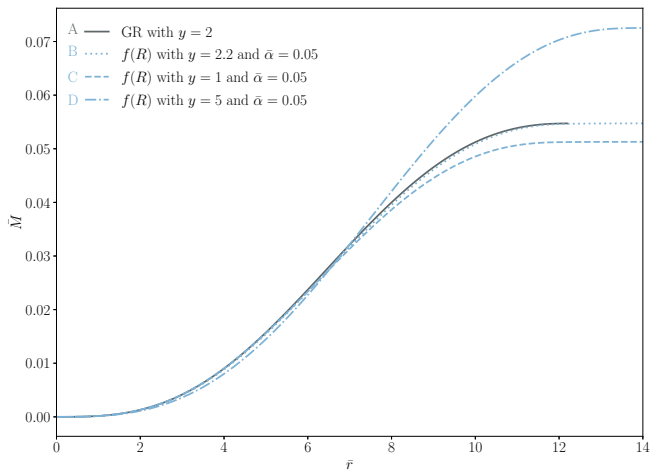}}
		\caption{ For every case the mass profiles correspond to $f(\text{R})$ theory are down to the mass profile obtained for the same case (same ``y'' value) but, using General Relativity. Is possible increase the mass obtained in R-squared in comparative we GR, if we choose for the first a higher value of ''y''. }\label{Fig04}
	\end{figure}

With this in mind, next we study the change in the maximum compactness of CDS as a function of the self-interacting strength constant $y$ value. The compactness of the CDS  in GR ($\bar \alpha=0$) are shown as  a solid black line in Fig.~\ref{compactness}. In GR, for $y=0$ (no self-interacting fermions) the maximum compactness is $C=0.11$, and it increases as $y$ increases.  To have an order of magnitude, let us recall that the Sun compactness is $\sim 10^{-5}$. Thus CDS can be very compact objects. As a comparison of other compact objects, in Fig.~\ref{compactness} we have included other important compactness values. For instance, the maximum compactness for a fluid star, which is given by the Buchdahl's limit and that is given by $C=4/9$~\cite{PhysRev.116.1027} is shown as a blue dotted line. In addition, the compactness of a Schwarzschild black hole, i.e. $C=1/2$, is shown as a green dashed line. Finally, the maximum compactness for a Boson Star which is given by $C=0.158$ ~\cite{AmaroSeoane:2010qx} is plotted as a red dotted line. In summary: Fermionic CDS have a maximum stable compactness bigger that Boson Stars. 

Now we can study Cold Dark Stars compactness in \mbox{R-square} gravity. It is important to recall that in \mbox{R-square} gravity one can distinguish between the asymptotic compactness $C = M_{ADM}/r_{*}$ - which captures contributions to the mass due to the non-vanishing value of the Ricci scalar. Nevertheless, $r_{*}$ is formally reached at infinity as in the case of Boson Stars, because the Ricci scalar decays as $\sim 1/r^2$ and thus $\text{R}=0$ can be reached only as $r_{*} \to \infty$. We used instead of the asymptotic compactness the following definition of compactness,
\begin{equation}
C = \frac{M_{99}}{R_{99}}\, ,\label{c99}
\end{equation}
where $M_{99}=0.99 \bar{M}_{ADM}$ and $R_{99}$ is the radius where $\bar{M}_{99}$ is reached.
 
We have plotted the maximum stable compactness (eq.~\eqref{c99}) in R-square gravity as a function of the self-interacting coupling $y$ in orange dashed lines of Fig.~\ref{compactness}. Two values of $\bar{\alpha}$ were chosen: $\bar{\alpha} = 0.05$ and $\bar{\alpha} = 1$. For increasing values of $\bar{\alpha}$, the maximum compactness is decreasing, because the maximum mass in \mbox{R-square} gravity decreases meanwhile $R_{99}$ is not severely affected. Note that there are configurations where the compactness for CDS in GR intersects the compactness for CDS in \mbox{R-square} gravity for different values of $y$. This intersection leads us to the confusion between dark matter properties with modifications of GR as we discuss in the next section.

\subsection{Confusing}
\begin{figure*}
	\centering
	\scalebox{0.45}{\input{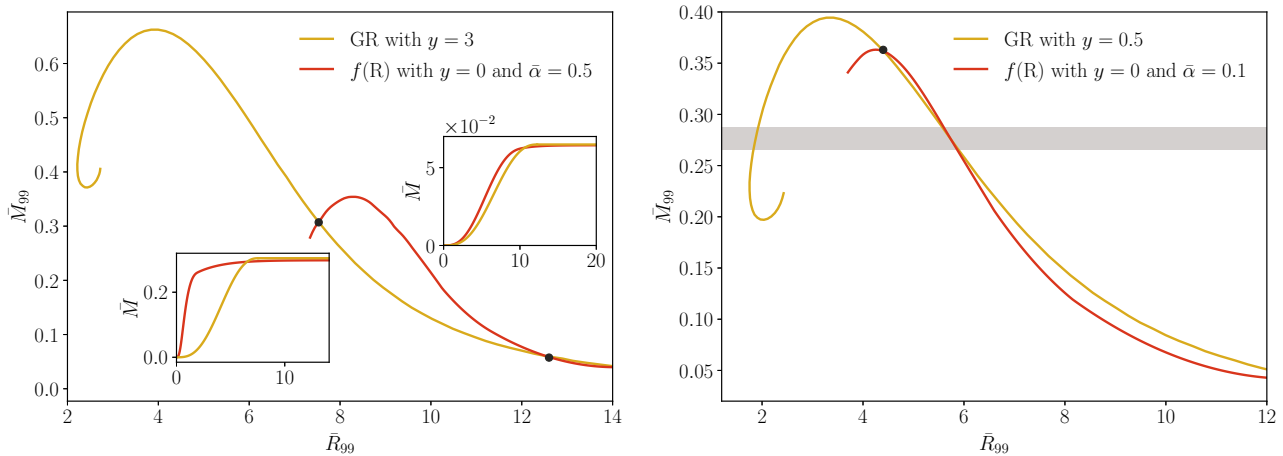}} 
\caption{The figures show the relationship between mass and radius in GR and $f(R)$. Left: The orange line correspond to all configurations (in the space parameter study) with GR and $y=3$. The red line represent the same, but in $f(R)$ with $y=0$. As see, exist confusing for some configurations (black points). This objects are indistinguishable one of other, because the have the same mass and radius. Right: Is illustrated the same idea, but whit other values for $y$ and $\bar{\alpha}$. Now given the typical astrophysical uncertainties, small variations on the compactness will be difficult to be disentangled, the GR and $f(R)$. Note that in this case there is an almost identical region (dark region) where  both plots intersects, and thus a infinite number of configurations can be confused}
\label{Fig1}
\end{figure*}

The previous results show the linear relationship between the free parameter, $y$, $\bar{\alpha}$ and the mass of the resulting CDS. In this section we will show that for a different selection of parameters $m$, $y$, $\bar{\alpha}$ and $\bar{p_0}$ there are equivalent configurations,  and thus, a confusing on the determination of the parameters is possible.

To illustrate this point, we have constructed three different configurations shown in Fig.~\ref{Fig04}
\begin{itemize}
\item Configuration A: Solid line corresponds to $\bar{\rho}_{0} = 4.1 \times 10^{-5}$, $y = 2$ (with self-interaction) $\bar{\alpha} = 0$ (General Relativity),
\item Configuration B: Dotted line corresponds to $\bar{\rho}_{0} = 4.1 \times 10^{-5}$, $y = 2.2$ (with self-interaction) $\bar{\alpha} = 0.05$ (R-square gravity),
\item Configuration C: Dotted line corresponds to $\bar{\rho}_{0} = 4.1 \times 10^{-5}$, $y=1$ (with self-interaction) $\bar{\alpha} = 0.05$ (R-square gravity),
\item Configuration D: Dashed line corresponds to $\bar{\rho}_{0} = 3.4 \times 10^{-5}$, $y=5$ (with self-interaction) $\bar{\alpha} = 0.05$ (R-square gravity).
\end{itemize}
Note that configuration B has the same central density as configuration A, but different interaction strength in R-square gravity. Nevertheless, the mass of the configuration and the radius are are almost indistinguishable. Thus, any test particle will follow the same trajectory in both configurations. 

On the other hand, Configuration C has a smaller central density, but due to the increase of the interaction strength $y$ the total mass increases and it is above the RG value. Thus, it seems that a there will be a value of $\rho_{0} \in [3,4]\times 10^{-5}$ and $y \in [1,5]$ for a R-square gravity with $\bar{\alpha} = 0.05$ will coincide with the GR case for $y=2$ and $\bar{\rho}_{0} = 4.1 \times 10^{-5}$. Thus any gravitational signal for this two objects will be almost identical for different values of the coupling constant.

There are other possible confusing possibilities in the determination of the parameters. Let us for instance construct all configurations with $y=3$ in GR and all configurations for $y=0$ in R-square gravity with $\bar{\alpha} = 0.5$. The plots $M_{99}$ {\it vs} $R_{99}$ for those configurations are shown in left panel of Fig.~\ref{Fig1}. Note that there are 2 points that intersect both plots. Those configurations corresponds to CDS with the same compactness. Thus, any other object will feel the same gravitational potential outside this CDS: no dynamical differences will be seen neither infill differences. Thus, if one luminous star is orbiting this cold Dark Star and this is the only observable we have to constraints the dark matter particle properties (remember that in this case that will be the mass of the dark matter particle and the self-interaction strength).

Finally, given the typical astrophysical uncertainties,  small variations on the compactness or the $M_{99}$ {\it vs} $R_{99}$ relation will be difficult to be disentangled. On the top of that, we can have the already discussed confusing between GR and R-square gravity. Let us illustrate this final possibility by constructing all possible configurations  with $y=0.5$ in GR and all configurations for $y=0$ in R-square gravity with $\bar \alpha=0.1$. The plots $M_{99}$ {\it vs} $R_{99}$ for those configurations are shown in right panel of Fig.~\ref{Fig1}. Note that in this case there is an almost identical region where both plots intersects, and thus a infinite number of configurations can be confused. 

\section{Conclusions}\label{sec:conclusions}
 
Modifications to General Relativity have been usually invoked to replace the role of dark matter or dark energy. Nevertheless, it could be possible that even with 
the existence of dark matter, there are possible modifications to GR. This modifications to GR seems to be mandatory specially if one looks for renormalizability of the gravitational interactions at high energies. Thus, we have consider a scenario where both dark matter and modifications to GR are present. In particular, we have consider dark matter as a self -interacting fermion in the degenerate limit and R-square gravity as a possible modification of GR. In this scenario, we have constructed self-gravitating structures made of this fermionic dark matter in both theories of gravity and studied their properties such as the compactness of the configurations. We have called this configurations Cold Dark Stars.
 
In the context of GR, we have shown that is possible to obtain configurations of CDS more compact that Boson Stars. Depending on the election of the mass of the fermion, CDS can be as massive and compact to be considered as black hole mimickers. 

We have shown that CDS in R-square gravity,  the bigger the quadratic term in the Einstein-Hilbert action, i.e. the bigger the value of $\bar{\alpha}$, the smaller the maximum mass of the resulting CDS configurations are obtained. This reduction in the mass will produce less compact CDS in R-square gravity in comparison with GR. 

Considering the possibility that dark matter might interact only through gravitational interactions, the determination of their properties will be only accessible by astrophysical observations. Thus, given that similar CDS can be obtained for different election of coupling constants $y$ or values of $\bar{\alpha}$ in R-square theories, thus, a possible confusion in the determination of the dark matter properties and possible modifications to general relativity could be possible.
 
\subsection*{Acknowledgements}
This work was partially support by CONACYT project  CB- 286651 and Conacyt-SNI. A.R. acknowledge support by CONACyT graduate scholarships No. 570326. We thank Alberto Diez-Tejedor and Gustavo Niz for very enlightening discussion regarding this work.

\subsection*{\label{sec:citeref} References}
\bibliographystyle{unsrt}

\bibliography{confusing.bib} 
\end{document}

%% file: RG.pgf
\begingroup%
\makeatletter%
\begin{pgfpicture}%
\pgfpathrectangle{\pgfpointorigin}{\pgfqpoint{10.293119in}{7.610251in}}%
\pgfusepath{use as bounding box, clip}%
\begin{pgfscope}%
\pgfsetbuttcap%
\pgfsetmiterjoin%
\definecolor{currentfill}{rgb}{1.000000,1.000000,1.000000}%
\pgfsetfillcolor{currentfill}%
\pgfsetlinewidth{0.000000pt}%
\definecolor{currentstroke}{rgb}{1.000000,1.000000,1.000000}%
\pgfsetstrokecolor{currentstroke}%
\pgfsetdash{}{0pt}%
\pgfpathmoveto{\pgfqpoint{0.000000in}{0.000000in}}%
\pgfpathlineto{\pgfqpoint{10.293119in}{0.000000in}}%
\pgfpathlineto{\pgfqpoint{10.293119in}{7.610251in}}%
\pgfpathlineto{\pgfqpoint{0.000000in}{7.610251in}}%
\pgfpathclose%
\pgfusepath{fill}%
\end{pgfscope}%
\begin{pgfscope}%
\pgfsetbuttcap%
\pgfsetmiterjoin%
\definecolor{currentfill}{rgb}{1.000000,1.000000,1.000000}%
\pgfsetfillcolor{currentfill}%
\pgfsetlinewidth{0.000000pt}%
\definecolor{currentstroke}{rgb}{0.000000,0.000000,0.000000}%
\pgfsetstrokecolor{currentstroke}%
\pgfsetstrokeopacity{0.000000}%
\pgfsetdash{}{0pt}%
\pgfpathmoveto{\pgfqpoint{0.783051in}{0.693364in}}%
\pgfpathlineto{\pgfqpoint{10.083051in}{0.693364in}}%
\pgfpathlineto{\pgfqpoint{10.083051in}{7.488364in}}%
\pgfpathlineto{\pgfqpoint{0.783051in}{7.488364in}}%
\pgfpathclose%
\pgfusepath{fill}%
\end{pgfscope}%
\begin{pgfscope}%
\pgfsetbuttcap%
\pgfsetroundjoin%
\definecolor{currentfill}{rgb}{0.000000,0.000000,0.000000}%
\pgfsetfillcolor{currentfill}%
\pgfsetlinewidth{0.803000pt}%
\definecolor{currentstroke}{rgb}{0.000000,0.000000,0.000000}%
\pgfsetstrokecolor{currentstroke}%
\pgfsetdash{}{0pt}%
\pgfsys@defobject{currentmarker}{\pgfqpoint{0.000000in}{-0.048611in}}{\pgfqpoint{0.000000in}{0.000000in}}{%
\pgfpathmoveto{\pgfqpoint{0.000000in}{0.000000in}}%
\pgfpathlineto{\pgfqpoint{0.000000in}{-0.048611in}}%
\pgfusepath{stroke,fill}%
}%
\begin{pgfscope}%
\pgfsys@transformshift{1.628506in}{0.693364in}%
\pgfsys@useobject{currentmarker}{}%
\end{pgfscope}%
\end{pgfscope}%
\begin{pgfscope}%
\definecolor{textcolor}{rgb}{0.000000,0.000000,0.000000}%
\pgfsetstrokecolor{textcolor}%
\pgfsetfillcolor{textcolor}%
\pgftext[x=1.628506in,y=0.596142in,,top]{\color{textcolor}\rmfamily\fontsize{16.000000}{19.200000}\selectfont \(\displaystyle 2\)}%
\end{pgfscope}%
\begin{pgfscope}%
\pgfsetbuttcap%
\pgfsetroundjoin%
\definecolor{currentfill}{rgb}{0.000000,0.000000,0.000000}%
\pgfsetfillcolor{currentfill}%
\pgfsetlinewidth{0.803000pt}%
\definecolor{currentstroke}{rgb}{0.000000,0.000000,0.000000}%
\pgfsetstrokecolor{currentstroke}%
\pgfsetdash{}{0pt}%
\pgfsys@defobject{currentmarker}{\pgfqpoint{0.000000in}{-0.048611in}}{\pgfqpoint{0.000000in}{0.000000in}}{%
\pgfpathmoveto{\pgfqpoint{0.000000in}{0.000000in}}%
\pgfpathlineto{\pgfqpoint{0.000000in}{-0.048611in}}%
\pgfusepath{stroke,fill}%
}%
\begin{pgfscope}%
\pgfsys@transformshift{3.319415in}{0.693364in}%
\pgfsys@useobject{currentmarker}{}%
\end{pgfscope}%
\end{pgfscope}%
\begin{pgfscope}%
\definecolor{textcolor}{rgb}{0.000000,0.000000,0.000000}%
\pgfsetstrokecolor{textcolor}%
\pgfsetfillcolor{textcolor}%
\pgftext[x=3.319415in,y=0.596142in,,top]{\color{textcolor}\rmfamily\fontsize{16.000000}{19.200000}\selectfont \(\displaystyle 4\)}%
\end{pgfscope}%
\begin{pgfscope}%
\pgfsetbuttcap%
\pgfsetroundjoin%
\definecolor{currentfill}{rgb}{0.000000,0.000000,0.000000}%
\pgfsetfillcolor{currentfill}%
\pgfsetlinewidth{0.803000pt}%
\definecolor{currentstroke}{rgb}{0.000000,0.000000,0.000000}%
\pgfsetstrokecolor{currentstroke}%
\pgfsetdash{}{0pt}%
\pgfsys@defobject{currentmarker}{\pgfqpoint{0.000000in}{-0.048611in}}{\pgfqpoint{0.000000in}{0.000000in}}{%
\pgfpathmoveto{\pgfqpoint{0.000000in}{0.000000in}}%
\pgfpathlineto{\pgfqpoint{0.000000in}{-0.048611in}}%
\pgfusepath{stroke,fill}%
}%
\begin{pgfscope}%
\pgfsys@transformshift{5.010324in}{0.693364in}%
\pgfsys@useobject{currentmarker}{}%
\end{pgfscope}%
\end{pgfscope}%
\begin{pgfscope}%
\definecolor{textcolor}{rgb}{0.000000,0.000000,0.000000}%
\pgfsetstrokecolor{textcolor}%
\pgfsetfillcolor{textcolor}%
\pgftext[x=5.010324in,y=0.596142in,,top]{\color{textcolor}\rmfamily\fontsize{16.000000}{19.200000}\selectfont \(\displaystyle 6\)}%
\end{pgfscope}%
\begin{pgfscope}%
\pgfsetbuttcap%
\pgfsetroundjoin%
\definecolor{currentfill}{rgb}{0.000000,0.000000,0.000000}%
\pgfsetfillcolor{currentfill}%
\pgfsetlinewidth{0.803000pt}%
\definecolor{currentstroke}{rgb}{0.000000,0.000000,0.000000}%
\pgfsetstrokecolor{currentstroke}%
\pgfsetdash{}{0pt}%
\pgfsys@defobject{currentmarker}{\pgfqpoint{0.000000in}{-0.048611in}}{\pgfqpoint{0.000000in}{0.000000in}}{%
\pgfpathmoveto{\pgfqpoint{0.000000in}{0.000000in}}%
\pgfpathlineto{\pgfqpoint{0.000000in}{-0.048611in}}%
\pgfusepath{stroke,fill}%
}%
\begin{pgfscope}%
\pgfsys@transformshift{6.701233in}{0.693364in}%
\pgfsys@useobject{currentmarker}{}%
\end{pgfscope}%
\end{pgfscope}%
\begin{pgfscope}%
\definecolor{textcolor}{rgb}{0.000000,0.000000,0.000000}%
\pgfsetstrokecolor{textcolor}%
\pgfsetfillcolor{textcolor}%
\pgftext[x=6.701233in,y=0.596142in,,top]{\color{textcolor}\rmfamily\fontsize{16.000000}{19.200000}\selectfont \(\displaystyle 8\)}%
\end{pgfscope}%
\begin{pgfscope}%
\pgfsetbuttcap%
\pgfsetroundjoin%
\definecolor{currentfill}{rgb}{0.000000,0.000000,0.000000}%
\pgfsetfillcolor{currentfill}%
\pgfsetlinewidth{0.803000pt}%
\definecolor{currentstroke}{rgb}{0.000000,0.000000,0.000000}%
\pgfsetstrokecolor{currentstroke}%
\pgfsetdash{}{0pt}%
\pgfsys@defobject{currentmarker}{\pgfqpoint{0.000000in}{-0.048611in}}{\pgfqpoint{0.000000in}{0.000000in}}{%
\pgfpathmoveto{\pgfqpoint{0.000000in}{0.000000in}}%
\pgfpathlineto{\pgfqpoint{0.000000in}{-0.048611in}}%
\pgfusepath{stroke,fill}%
}%
\begin{pgfscope}%
\pgfsys@transformshift{8.392142in}{0.693364in}%
\pgfsys@useobject{currentmarker}{}%
\end{pgfscope}%
\end{pgfscope}%
\begin{pgfscope}%
\definecolor{textcolor}{rgb}{0.000000,0.000000,0.000000}%
\pgfsetstrokecolor{textcolor}%
\pgfsetfillcolor{textcolor}%
\pgftext[x=8.392142in,y=0.596142in,,top]{\color{textcolor}\rmfamily\fontsize{16.000000}{19.200000}\selectfont \(\displaystyle 10\)}%
\end{pgfscope}%
\begin{pgfscope}%
\pgfsetbuttcap%
\pgfsetroundjoin%
\definecolor{currentfill}{rgb}{0.000000,0.000000,0.000000}%
\pgfsetfillcolor{currentfill}%
\pgfsetlinewidth{0.803000pt}%
\definecolor{currentstroke}{rgb}{0.000000,0.000000,0.000000}%
\pgfsetstrokecolor{currentstroke}%
\pgfsetdash{}{0pt}%
\pgfsys@defobject{currentmarker}{\pgfqpoint{0.000000in}{-0.048611in}}{\pgfqpoint{0.000000in}{0.000000in}}{%
\pgfpathmoveto{\pgfqpoint{0.000000in}{0.000000in}}%
\pgfpathlineto{\pgfqpoint{0.000000in}{-0.048611in}}%
\pgfusepath{stroke,fill}%
}%
\begin{pgfscope}%
\pgfsys@transformshift{10.083051in}{0.693364in}%
\pgfsys@useobject{currentmarker}{}%
\end{pgfscope}%
\end{pgfscope}%
\begin{pgfscope}%
\definecolor{textcolor}{rgb}{0.000000,0.000000,0.000000}%
\pgfsetstrokecolor{textcolor}%
\pgfsetfillcolor{textcolor}%
\pgftext[x=10.083051in,y=0.596142in,,top]{\color{textcolor}\rmfamily\fontsize{16.000000}{19.200000}\selectfont \(\displaystyle 12\)}%
\end{pgfscope}%
\begin{pgfscope}%
\definecolor{textcolor}{rgb}{0.000000,0.000000,0.000000}%
\pgfsetstrokecolor{textcolor}%
\pgfsetfillcolor{textcolor}%
\pgftext[x=5.433051in,y=0.313349in,,top]{\color{textcolor}\rmfamily\fontsize{18.000000}{21.600000}\selectfont \(\displaystyle \bar{R}\)}%
\end{pgfscope}%
\begin{pgfscope}%
\pgfsetbuttcap%
\pgfsetroundjoin%
\definecolor{currentfill}{rgb}{0.000000,0.000000,0.000000}%
\pgfsetfillcolor{currentfill}%
\pgfsetlinewidth{0.803000pt}%
\definecolor{currentstroke}{rgb}{0.000000,0.000000,0.000000}%
\pgfsetstrokecolor{currentstroke}%
\pgfsetdash{}{0pt}%
\pgfsys@defobject{currentmarker}{\pgfqpoint{-0.048611in}{0.000000in}}{\pgfqpoint{0.000000in}{0.000000in}}{%
\pgfpathmoveto{\pgfqpoint{0.000000in}{0.000000in}}%
\pgfpathlineto{\pgfqpoint{-0.048611in}{0.000000in}}%
\pgfusepath{stroke,fill}%
}%
\begin{pgfscope}%
\pgfsys@transformshift{0.783051in}{0.941978in}%
\pgfsys@useobject{currentmarker}{}%
\end{pgfscope}%
\end{pgfscope}%
\begin{pgfscope}%
\definecolor{textcolor}{rgb}{0.000000,0.000000,0.000000}%
\pgfsetstrokecolor{textcolor}%
\pgfsetfillcolor{textcolor}%
\pgftext[x=0.400416in,y=0.858644in,left,base]{\color{textcolor}\rmfamily\fontsize{16.000000}{19.200000}\selectfont \(\displaystyle 0.0\)}%
\end{pgfscope}%
\begin{pgfscope}%
\pgfsetbuttcap%
\pgfsetroundjoin%
\definecolor{currentfill}{rgb}{0.000000,0.000000,0.000000}%
\pgfsetfillcolor{currentfill}%
\pgfsetlinewidth{0.803000pt}%
\definecolor{currentstroke}{rgb}{0.000000,0.000000,0.000000}%
\pgfsetstrokecolor{currentstroke}%
\pgfsetdash{}{0pt}%
\pgfsys@defobject{currentmarker}{\pgfqpoint{-0.048611in}{0.000000in}}{\pgfqpoint{0.000000in}{0.000000in}}{%
\pgfpathmoveto{\pgfqpoint{0.000000in}{0.000000in}}%
\pgfpathlineto{\pgfqpoint{-0.048611in}{0.000000in}}%
\pgfusepath{stroke,fill}%
}%
\begin{pgfscope}%
\pgfsys@transformshift{0.783051in}{2.238966in}%
\pgfsys@useobject{currentmarker}{}%
\end{pgfscope}%
\end{pgfscope}%
\begin{pgfscope}%
\definecolor{textcolor}{rgb}{0.000000,0.000000,0.000000}%
\pgfsetstrokecolor{textcolor}%
\pgfsetfillcolor{textcolor}%
\pgftext[x=0.400416in,y=2.155632in,left,base]{\color{textcolor}\rmfamily\fontsize{16.000000}{19.200000}\selectfont \(\displaystyle 0.2\)}%
\end{pgfscope}%
\begin{pgfscope}%
\pgfsetbuttcap%
\pgfsetroundjoin%
\definecolor{currentfill}{rgb}{0.000000,0.000000,0.000000}%
\pgfsetfillcolor{currentfill}%
\pgfsetlinewidth{0.803000pt}%
\definecolor{currentstroke}{rgb}{0.000000,0.000000,0.000000}%
\pgfsetstrokecolor{currentstroke}%
\pgfsetdash{}{0pt}%
\pgfsys@defobject{currentmarker}{\pgfqpoint{-0.048611in}{0.000000in}}{\pgfqpoint{0.000000in}{0.000000in}}{%
\pgfpathmoveto{\pgfqpoint{0.000000in}{0.000000in}}%
\pgfpathlineto{\pgfqpoint{-0.048611in}{0.000000in}}%
\pgfusepath{stroke,fill}%
}%
\begin{pgfscope}%
\pgfsys@transformshift{0.783051in}{3.535954in}%
\pgfsys@useobject{currentmarker}{}%
\end{pgfscope}%
\end{pgfscope}%
\begin{pgfscope}%
\definecolor{textcolor}{rgb}{0.000000,0.000000,0.000000}%
\pgfsetstrokecolor{textcolor}%
\pgfsetfillcolor{textcolor}%
\pgftext[x=0.400416in,y=3.452620in,left,base]{\color{textcolor}\rmfamily\fontsize{16.000000}{19.200000}\selectfont \(\displaystyle 0.4\)}%
\end{pgfscope}%
\begin{pgfscope}%
\pgfsetbuttcap%
\pgfsetroundjoin%
\definecolor{currentfill}{rgb}{0.000000,0.000000,0.000000}%
\pgfsetfillcolor{currentfill}%
\pgfsetlinewidth{0.803000pt}%
\definecolor{currentstroke}{rgb}{0.000000,0.000000,0.000000}%
\pgfsetstrokecolor{currentstroke}%
\pgfsetdash{}{0pt}%
\pgfsys@defobject{currentmarker}{\pgfqpoint{-0.048611in}{0.000000in}}{\pgfqpoint{0.000000in}{0.000000in}}{%
\pgfpathmoveto{\pgfqpoint{0.000000in}{0.000000in}}%
\pgfpathlineto{\pgfqpoint{-0.048611in}{0.000000in}}%
\pgfusepath{stroke,fill}%
}%
\begin{pgfscope}%
\pgfsys@transformshift{0.783051in}{4.832942in}%
\pgfsys@useobject{currentmarker}{}%
\end{pgfscope}%
\end{pgfscope}%
\begin{pgfscope}%
\definecolor{textcolor}{rgb}{0.000000,0.000000,0.000000}%
\pgfsetstrokecolor{textcolor}%
\pgfsetfillcolor{textcolor}%
\pgftext[x=0.400416in,y=4.749609in,left,base]{\color{textcolor}\rmfamily\fontsize{16.000000}{19.200000}\selectfont \(\displaystyle 0.6\)}%
\end{pgfscope}%
\begin{pgfscope}%
\pgfsetbuttcap%
\pgfsetroundjoin%
\definecolor{currentfill}{rgb}{0.000000,0.000000,0.000000}%
\pgfsetfillcolor{currentfill}%
\pgfsetlinewidth{0.803000pt}%
\definecolor{currentstroke}{rgb}{0.000000,0.000000,0.000000}%
\pgfsetstrokecolor{currentstroke}%
\pgfsetdash{}{0pt}%
\pgfsys@defobject{currentmarker}{\pgfqpoint{-0.048611in}{0.000000in}}{\pgfqpoint{0.000000in}{0.000000in}}{%
\pgfpathmoveto{\pgfqpoint{0.000000in}{0.000000in}}%
\pgfpathlineto{\pgfqpoint{-0.048611in}{0.000000in}}%
\pgfusepath{stroke,fill}%
}%
\begin{pgfscope}%
\pgfsys@transformshift{0.783051in}{6.129930in}%
\pgfsys@useobject{currentmarker}{}%
\end{pgfscope}%
\end{pgfscope}%
\begin{pgfscope}%
\definecolor{textcolor}{rgb}{0.000000,0.000000,0.000000}%
\pgfsetstrokecolor{textcolor}%
\pgfsetfillcolor{textcolor}%
\pgftext[x=0.400416in,y=6.046597in,left,base]{\color{textcolor}\rmfamily\fontsize{16.000000}{19.200000}\selectfont \(\displaystyle 0.8\)}%
\end{pgfscope}%
\begin{pgfscope}%
\pgfsetbuttcap%
\pgfsetroundjoin%
\definecolor{currentfill}{rgb}{0.000000,0.000000,0.000000}%
\pgfsetfillcolor{currentfill}%
\pgfsetlinewidth{0.803000pt}%
\definecolor{currentstroke}{rgb}{0.000000,0.000000,0.000000}%
\pgfsetstrokecolor{currentstroke}%
\pgfsetdash{}{0pt}%
\pgfsys@defobject{currentmarker}{\pgfqpoint{-0.048611in}{0.000000in}}{\pgfqpoint{0.000000in}{0.000000in}}{%
\pgfpathmoveto{\pgfqpoint{0.000000in}{0.000000in}}%
\pgfpathlineto{\pgfqpoint{-0.048611in}{0.000000in}}%
\pgfusepath{stroke,fill}%
}%
\begin{pgfscope}%
\pgfsys@transformshift{0.783051in}{7.426918in}%
\pgfsys@useobject{currentmarker}{}%
\end{pgfscope}%
\end{pgfscope}%
\begin{pgfscope}%
\definecolor{textcolor}{rgb}{0.000000,0.000000,0.000000}%
\pgfsetstrokecolor{textcolor}%
\pgfsetfillcolor{textcolor}%
\pgftext[x=0.400416in,y=7.343585in,left,base]{\color{textcolor}\rmfamily\fontsize{16.000000}{19.200000}\selectfont \(\displaystyle 1.0\)}%
\end{pgfscope}%
\begin{pgfscope}%
\definecolor{textcolor}{rgb}{0.000000,0.000000,0.000000}%
\pgfsetstrokecolor{textcolor}%
\pgfsetfillcolor{textcolor}%
\pgftext[x=0.330971in,y=4.090864in,,bottom,rotate=90.000000]{\color{textcolor}\rmfamily\fontsize{18.000000}{21.600000}\selectfont \(\displaystyle \bar{M}_{\mathrm{ADM}}\)}%
\end{pgfscope}%
\begin{pgfscope}%
\pgfpathrectangle{\pgfqpoint{0.783051in}{0.693364in}}{\pgfqpoint{9.300000in}{6.795000in}}%
\pgfusepath{clip}%
\pgfsetrectcap%
\pgfsetroundjoin%
\pgfsetlinewidth{2.007500pt}%
\definecolor{currentstroke}{rgb}{0.149020,0.137255,0.133333}%
\pgfsetstrokecolor{currentstroke}%
\pgfsetdash{}{0pt}%
\pgfpathmoveto{\pgfqpoint{10.084051in}{1.269894in}}%
\pgfpathlineto{\pgfqpoint{9.999681in}{1.277695in}}%
\pgfpathlineto{\pgfqpoint{9.755524in}{1.302022in}}%
\pgfpathlineto{\pgfqpoint{9.505565in}{1.327999in}}%
\pgfpathlineto{\pgfqpoint{9.266261in}{1.355767in}}%
\pgfpathlineto{\pgfqpoint{9.045644in}{1.385356in}}%
\pgfpathlineto{\pgfqpoint{8.806488in}{1.416917in}}%
\pgfpathlineto{\pgfqpoint{8.589347in}{1.450514in}}%
\pgfpathlineto{\pgfqpoint{8.387404in}{1.486271in}}%
\pgfpathlineto{\pgfqpoint{8.151780in}{1.524290in}}%
\pgfpathlineto{\pgfqpoint{7.959480in}{1.564645in}}%
\pgfpathlineto{\pgfqpoint{7.738460in}{1.607439in}}%
\pgfpathlineto{\pgfqpoint{7.545429in}{1.652756in}}%
\pgfpathlineto{\pgfqpoint{7.339860in}{1.700669in}}%
\pgfpathlineto{\pgfqpoint{7.153361in}{1.751252in}}%
\pgfpathlineto{\pgfqpoint{6.956407in}{1.804536in}}%
\pgfpathlineto{\pgfqpoint{6.769169in}{1.860553in}}%
\pgfpathlineto{\pgfqpoint{6.586114in}{1.919322in}}%
\pgfpathlineto{\pgfqpoint{6.404127in}{1.980862in}}%
\pgfpathlineto{\pgfqpoint{6.228736in}{2.045049in}}%
\pgfpathlineto{\pgfqpoint{6.058116in}{2.111832in}}%
\pgfpathlineto{\pgfqpoint{5.884328in}{2.181098in}}%
\pgfpathlineto{\pgfqpoint{5.710372in}{2.252652in}}%
\pgfpathlineto{\pgfqpoint{5.540200in}{2.326310in}}%
\pgfpathlineto{\pgfqpoint{5.381346in}{2.401756in}}%
\pgfpathlineto{\pgfqpoint{5.217938in}{2.478688in}}%
\pgfpathlineto{\pgfqpoint{5.054685in}{2.556699in}}%
\pgfpathlineto{\pgfqpoint{4.902620in}{2.635320in}}%
\pgfpathlineto{\pgfqpoint{4.746594in}{2.714072in}}%
\pgfpathlineto{\pgfqpoint{4.590285in}{2.792315in}}%
\pgfpathlineto{\pgfqpoint{4.435235in}{2.869406in}}%
\pgfpathlineto{\pgfqpoint{4.287121in}{2.944662in}}%
\pgfpathlineto{\pgfqpoint{4.142374in}{3.017335in}}%
\pgfpathlineto{\pgfqpoint{3.993230in}{3.086607in}}%
\pgfpathlineto{\pgfqpoint{3.853207in}{3.151692in}}%
\pgfpathlineto{\pgfqpoint{3.712706in}{3.211780in}}%
\pgfpathlineto{\pgfqpoint{3.579248in}{3.266086in}}%
\pgfpathlineto{\pgfqpoint{3.438623in}{3.313841in}}%
\pgfpathlineto{\pgfqpoint{3.303143in}{3.354381in}}%
\pgfpathlineto{\pgfqpoint{3.174962in}{3.387061in}}%
\pgfpathlineto{\pgfqpoint{3.047710in}{3.411413in}}%
\pgfpathlineto{\pgfqpoint{2.920551in}{3.427031in}}%
\pgfpathlineto{\pgfqpoint{2.800613in}{3.433655in}}%
\pgfpathlineto{\pgfqpoint{2.687000in}{3.431215in}}%
\pgfpathlineto{\pgfqpoint{2.570359in}{3.419761in}}%
\pgfpathlineto{\pgfqpoint{2.462928in}{3.399513in}}%
\pgfpathlineto{\pgfqpoint{2.357523in}{3.370823in}}%
\pgfpathlineto{\pgfqpoint{2.257767in}{3.334217in}}%
\pgfpathlineto{\pgfqpoint{2.166760in}{3.290329in}}%
\pgfpathlineto{\pgfqpoint{2.076419in}{3.239921in}}%
\pgfpathlineto{\pgfqpoint{1.990209in}{3.183813in}}%
\pgfpathlineto{\pgfqpoint{1.910042in}{3.122910in}}%
\pgfpathlineto{\pgfqpoint{1.834398in}{3.058188in}}%
\pgfpathlineto{\pgfqpoint{1.767175in}{2.990588in}}%
\pgfpathlineto{\pgfqpoint{1.704606in}{2.921074in}}%
\pgfpathlineto{\pgfqpoint{1.650856in}{2.850617in}}%
\pgfpathlineto{\pgfqpoint{1.603101in}{2.780147in}}%
\pgfpathlineto{\pgfqpoint{1.564171in}{2.710587in}}%
\pgfpathlineto{\pgfqpoint{1.524958in}{2.642749in}}%
\pgfpathlineto{\pgfqpoint{1.497855in}{2.577540in}}%
\pgfpathlineto{\pgfqpoint{1.478781in}{2.515683in}}%
\pgfpathlineto{\pgfqpoint{1.465517in}{2.457985in}}%
\pgfpathlineto{\pgfqpoint{1.463174in}{2.405188in}}%
\pgfpathlineto{\pgfqpoint{1.470710in}{2.358087in}}%
\pgfpathlineto{\pgfqpoint{1.490162in}{2.317395in}}%
\pgfpathlineto{\pgfqpoint{1.514456in}{2.283886in}}%
\pgfpathlineto{\pgfqpoint{1.553312in}{2.258259in}}%
\pgfpathlineto{\pgfqpoint{1.598580in}{2.241116in}}%
\pgfpathlineto{\pgfqpoint{1.652142in}{2.232853in}}%
\pgfpathlineto{\pgfqpoint{1.707792in}{2.233572in}}%
\pgfpathlineto{\pgfqpoint{1.772694in}{2.242890in}}%
\pgfpathlineto{\pgfqpoint{1.820725in}{2.259849in}}%
\pgfpathlineto{\pgfqpoint{1.872719in}{2.282942in}}%
\pgfpathlineto{\pgfqpoint{1.912235in}{2.310333in}}%
\pgfpathlineto{\pgfqpoint{1.946115in}{2.339910in}}%
\pgfpathlineto{\pgfqpoint{1.966816in}{2.369654in}}%
\pgfpathlineto{\pgfqpoint{1.975426in}{2.397870in}}%
\pgfpathlineto{\pgfqpoint{1.978045in}{2.423187in}}%
\pgfusepath{stroke}%
\end{pgfscope}%
\begin{pgfscope}%
\pgfpathrectangle{\pgfqpoint{0.783051in}{0.693364in}}{\pgfqpoint{9.300000in}{6.795000in}}%
\pgfusepath{clip}%
\pgfsetbuttcap%
\pgfsetroundjoin%
\pgfsetlinewidth{2.007500pt}%
\definecolor{currentstroke}{rgb}{0.160784,0.329412,0.549020}%
\pgfsetstrokecolor{currentstroke}%
\pgfsetdash{{7.400000pt}{3.200000pt}}{0.000000pt}%
\pgfpathmoveto{\pgfqpoint{10.084051in}{1.281881in}}%
\pgfpathlineto{\pgfqpoint{10.063699in}{1.283757in}}%
\pgfpathlineto{\pgfqpoint{9.819235in}{1.308833in}}%
\pgfpathlineto{\pgfqpoint{9.588127in}{1.335662in}}%
\pgfpathlineto{\pgfqpoint{9.342208in}{1.364386in}}%
\pgfpathlineto{\pgfqpoint{9.101331in}{1.395051in}}%
\pgfpathlineto{\pgfqpoint{8.874868in}{1.427785in}}%
\pgfpathlineto{\pgfqpoint{8.657374in}{1.462725in}}%
\pgfpathlineto{\pgfqpoint{8.438759in}{1.499962in}}%
\pgfpathlineto{\pgfqpoint{8.238807in}{1.539635in}}%
\pgfpathlineto{\pgfqpoint{8.021709in}{1.581841in}}%
\pgfpathlineto{\pgfqpoint{7.839913in}{1.626706in}}%
\pgfpathlineto{\pgfqpoint{7.621925in}{1.674313in}}%
\pgfpathlineto{\pgfqpoint{7.421299in}{1.724773in}}%
\pgfpathlineto{\pgfqpoint{7.232243in}{1.778175in}}%
\pgfpathlineto{\pgfqpoint{7.050569in}{1.834580in}}%
\pgfpathlineto{\pgfqpoint{6.853877in}{1.894050in}}%
\pgfpathlineto{\pgfqpoint{6.676424in}{1.956617in}}%
\pgfpathlineto{\pgfqpoint{6.501842in}{2.022337in}}%
\pgfpathlineto{\pgfqpoint{6.319707in}{2.091110in}}%
\pgfpathlineto{\pgfqpoint{6.143898in}{2.162909in}}%
\pgfpathlineto{\pgfqpoint{5.971811in}{2.237645in}}%
\pgfpathlineto{\pgfqpoint{5.802184in}{2.315165in}}%
\pgfpathlineto{\pgfqpoint{5.636737in}{2.395266in}}%
\pgfpathlineto{\pgfqpoint{5.472928in}{2.477682in}}%
\pgfpathlineto{\pgfqpoint{5.311919in}{2.562104in}}%
\pgfpathlineto{\pgfqpoint{5.151514in}{2.648126in}}%
\pgfpathlineto{\pgfqpoint{4.997477in}{2.735307in}}%
\pgfpathlineto{\pgfqpoint{4.845786in}{2.823097in}}%
\pgfpathlineto{\pgfqpoint{4.685373in}{2.910880in}}%
\pgfpathlineto{\pgfqpoint{4.537351in}{2.997970in}}%
\pgfpathlineto{\pgfqpoint{4.387307in}{3.083621in}}%
\pgfpathlineto{\pgfqpoint{4.236053in}{3.167055in}}%
\pgfpathlineto{\pgfqpoint{4.090597in}{3.247358in}}%
\pgfpathlineto{\pgfqpoint{3.952442in}{3.323664in}}%
\pgfpathlineto{\pgfqpoint{3.805186in}{3.395079in}}%
\pgfpathlineto{\pgfqpoint{3.665605in}{3.460699in}}%
\pgfpathlineto{\pgfqpoint{3.531307in}{3.519652in}}%
\pgfpathlineto{\pgfqpoint{3.397034in}{3.571140in}}%
\pgfpathlineto{\pgfqpoint{3.270236in}{3.614427in}}%
\pgfpathlineto{\pgfqpoint{3.133668in}{3.648863in}}%
\pgfpathlineto{\pgfqpoint{3.005936in}{3.673969in}}%
\pgfpathlineto{\pgfqpoint{2.883360in}{3.689374in}}%
\pgfpathlineto{\pgfqpoint{2.765183in}{3.694882in}}%
\pgfpathlineto{\pgfqpoint{2.646688in}{3.690468in}}%
\pgfpathlineto{\pgfqpoint{2.533911in}{3.676268in}}%
\pgfpathlineto{\pgfqpoint{2.425280in}{3.652574in}}%
\pgfpathlineto{\pgfqpoint{2.324440in}{3.619887in}}%
\pgfpathlineto{\pgfqpoint{2.223689in}{3.578816in}}%
\pgfpathlineto{\pgfqpoint{2.126727in}{3.530072in}}%
\pgfpathlineto{\pgfqpoint{2.039267in}{3.474523in}}%
\pgfpathlineto{\pgfqpoint{1.953447in}{3.413045in}}%
\pgfpathlineto{\pgfqpoint{1.871324in}{3.346633in}}%
\pgfpathlineto{\pgfqpoint{1.797521in}{3.276260in}}%
\pgfpathlineto{\pgfqpoint{1.726943in}{3.202943in}}%
\pgfpathlineto{\pgfqpoint{1.663956in}{3.127636in}}%
\pgfpathlineto{\pgfqpoint{1.606759in}{3.051293in}}%
\pgfpathlineto{\pgfqpoint{1.556362in}{2.974834in}}%
\pgfpathlineto{\pgfqpoint{1.506661in}{2.899139in}}%
\pgfpathlineto{\pgfqpoint{1.468667in}{2.824948in}}%
\pgfpathlineto{\pgfqpoint{1.433516in}{2.753097in}}%
\pgfpathlineto{\pgfqpoint{1.407488in}{2.684239in}}%
\pgfpathlineto{\pgfqpoint{1.387956in}{2.619049in}}%
\pgfpathlineto{\pgfqpoint{1.378757in}{2.558217in}}%
\pgfpathlineto{\pgfqpoint{1.375193in}{2.502379in}}%
\pgfpathlineto{\pgfqpoint{1.380794in}{2.452154in}}%
\pgfpathlineto{\pgfqpoint{1.395542in}{2.408210in}}%
\pgfpathlineto{\pgfqpoint{1.416589in}{2.371277in}}%
\pgfpathlineto{\pgfqpoint{1.450843in}{2.341957in}}%
\pgfpathlineto{\pgfqpoint{1.490470in}{2.320891in}}%
\pgfpathlineto{\pgfqpoint{1.540353in}{2.308642in}}%
\pgfpathlineto{\pgfqpoint{1.595937in}{2.305451in}}%
\pgfpathlineto{\pgfqpoint{1.654192in}{2.311312in}}%
\pgfpathlineto{\pgfqpoint{1.712897in}{2.325587in}}%
\pgfpathlineto{\pgfqpoint{1.770591in}{2.347163in}}%
\pgfpathlineto{\pgfqpoint{1.823306in}{2.374363in}}%
\pgfpathlineto{\pgfqpoint{1.863544in}{2.405158in}}%
\pgfpathlineto{\pgfqpoint{1.889021in}{2.437399in}}%
\pgfusepath{stroke}%
\end{pgfscope}%
\begin{pgfscope}%
\pgfpathrectangle{\pgfqpoint{0.783051in}{0.693364in}}{\pgfqpoint{9.300000in}{6.795000in}}%
\pgfusepath{clip}%
\pgfsetbuttcap%
\pgfsetroundjoin%
\pgfsetlinewidth{2.007500pt}%
\definecolor{currentstroke}{rgb}{0.160784,0.329412,0.549020}%
\pgfsetstrokecolor{currentstroke}%
\pgfsetdash{{7.400000pt}{3.200000pt}}{0.000000pt}%
\pgfpathmoveto{\pgfqpoint{10.084051in}{1.319592in}}%
\pgfpathlineto{\pgfqpoint{9.999787in}{1.329162in}}%
\pgfpathlineto{\pgfqpoint{9.765896in}{1.358514in}}%
\pgfpathlineto{\pgfqpoint{9.554548in}{1.390054in}}%
\pgfpathlineto{\pgfqpoint{9.306076in}{1.423873in}}%
\pgfpathlineto{\pgfqpoint{9.090217in}{1.460150in}}%
\pgfpathlineto{\pgfqpoint{8.866710in}{1.499014in}}%
\pgfpathlineto{\pgfqpoint{8.665655in}{1.540677in}}%
\pgfpathlineto{\pgfqpoint{8.455107in}{1.585262in}}%
\pgfpathlineto{\pgfqpoint{8.250981in}{1.632922in}}%
\pgfpathlineto{\pgfqpoint{8.049546in}{1.683869in}}%
\pgfpathlineto{\pgfqpoint{7.868677in}{1.738231in}}%
\pgfpathlineto{\pgfqpoint{7.657411in}{1.796175in}}%
\pgfpathlineto{\pgfqpoint{7.468464in}{1.857864in}}%
\pgfpathlineto{\pgfqpoint{7.291730in}{1.923421in}}%
\pgfpathlineto{\pgfqpoint{7.119687in}{1.992976in}}%
\pgfpathlineto{\pgfqpoint{6.930499in}{2.066642in}}%
\pgfpathlineto{\pgfqpoint{6.753572in}{2.144497in}}%
\pgfpathlineto{\pgfqpoint{6.580978in}{2.226613in}}%
\pgfpathlineto{\pgfqpoint{6.410009in}{2.312927in}}%
\pgfpathlineto{\pgfqpoint{6.238754in}{2.403439in}}%
\pgfpathlineto{\pgfqpoint{6.071246in}{2.498041in}}%
\pgfpathlineto{\pgfqpoint{5.908084in}{2.596561in}}%
\pgfpathlineto{\pgfqpoint{5.751482in}{2.698760in}}%
\pgfpathlineto{\pgfqpoint{5.587414in}{2.804325in}}%
\pgfpathlineto{\pgfqpoint{5.431329in}{2.912820in}}%
\pgfpathlineto{\pgfqpoint{5.278734in}{3.023744in}}%
\pgfpathlineto{\pgfqpoint{5.118917in}{3.136462in}}%
\pgfpathlineto{\pgfqpoint{4.964718in}{3.250256in}}%
\pgfpathlineto{\pgfqpoint{4.813978in}{3.364260in}}%
\pgfpathlineto{\pgfqpoint{4.662733in}{3.477549in}}%
\pgfpathlineto{\pgfqpoint{4.516612in}{3.589067in}}%
\pgfpathlineto{\pgfqpoint{4.366623in}{3.697708in}}%
\pgfpathlineto{\pgfqpoint{4.219581in}{3.802230in}}%
\pgfpathlineto{\pgfqpoint{4.075136in}{3.901392in}}%
\pgfpathlineto{\pgfqpoint{3.934671in}{3.993960in}}%
\pgfpathlineto{\pgfqpoint{3.793661in}{4.078662in}}%
\pgfpathlineto{\pgfqpoint{3.654286in}{4.154322in}}%
\pgfpathlineto{\pgfqpoint{3.514323in}{4.219840in}}%
\pgfpathlineto{\pgfqpoint{3.379572in}{4.274243in}}%
\pgfpathlineto{\pgfqpoint{3.249732in}{4.316704in}}%
\pgfpathlineto{\pgfqpoint{3.117498in}{4.346628in}}%
\pgfpathlineto{\pgfqpoint{2.995932in}{4.363614in}}%
\pgfpathlineto{\pgfqpoint{2.869882in}{4.367507in}}%
\pgfpathlineto{\pgfqpoint{2.749276in}{4.358370in}}%
\pgfpathlineto{\pgfqpoint{2.636033in}{4.336571in}}%
\pgfpathlineto{\pgfqpoint{2.528297in}{4.302636in}}%
\pgfpathlineto{\pgfqpoint{2.415885in}{4.257361in}}%
\pgfpathlineto{\pgfqpoint{2.314583in}{4.201683in}}%
\pgfpathlineto{\pgfqpoint{2.216852in}{4.136726in}}%
\pgfpathlineto{\pgfqpoint{2.125346in}{4.063704in}}%
\pgfpathlineto{\pgfqpoint{2.039341in}{3.983935in}}%
\pgfpathlineto{\pgfqpoint{1.962919in}{3.898751in}}%
\pgfpathlineto{\pgfqpoint{1.885644in}{3.809505in}}%
\pgfpathlineto{\pgfqpoint{1.817146in}{3.717574in}}%
\pgfpathlineto{\pgfqpoint{1.753516in}{3.624241in}}%
\pgfpathlineto{\pgfqpoint{1.698012in}{3.530785in}}%
\pgfpathlineto{\pgfqpoint{1.649154in}{3.438362in}}%
\pgfpathlineto{\pgfqpoint{1.607532in}{3.348131in}}%
\pgfpathlineto{\pgfqpoint{1.574907in}{3.261134in}}%
\pgfpathlineto{\pgfqpoint{1.546522in}{3.178396in}}%
\pgfpathlineto{\pgfqpoint{1.530161in}{3.100808in}}%
\pgfpathlineto{\pgfqpoint{1.520162in}{3.029381in}}%
\pgfpathlineto{\pgfqpoint{1.516641in}{2.964905in}}%
\pgfpathlineto{\pgfqpoint{1.526736in}{2.908334in}}%
\pgfpathlineto{\pgfqpoint{1.543209in}{2.860418in}}%
\pgfpathlineto{\pgfqpoint{1.566618in}{2.821947in}}%
\pgfpathlineto{\pgfqpoint{1.600215in}{2.793575in}}%
\pgfpathlineto{\pgfqpoint{1.643117in}{2.775723in}}%
\pgfpathlineto{\pgfqpoint{1.688865in}{2.768487in}}%
\pgfpathlineto{\pgfqpoint{1.738027in}{2.771564in}}%
\pgfpathlineto{\pgfqpoint{1.783886in}{2.784014in}}%
\pgfpathlineto{\pgfqpoint{1.831409in}{2.804366in}}%
\pgfpathlineto{\pgfqpoint{1.869750in}{2.830666in}}%
\pgfpathlineto{\pgfqpoint{1.903569in}{2.860680in}}%
\pgfpathlineto{\pgfqpoint{1.931879in}{2.892220in}}%
\pgfpathlineto{\pgfqpoint{1.949395in}{2.923173in}}%
\pgfusepath{stroke}%
\end{pgfscope}%
\begin{pgfscope}%
\pgfpathrectangle{\pgfqpoint{0.783051in}{0.693364in}}{\pgfqpoint{9.300000in}{6.795000in}}%
\pgfusepath{clip}%
\pgfsetbuttcap%
\pgfsetroundjoin%
\pgfsetlinewidth{2.007500pt}%
\definecolor{currentstroke}{rgb}{0.160784,0.329412,0.549020}%
\pgfsetstrokecolor{currentstroke}%
\pgfsetdash{{7.400000pt}{3.200000pt}}{0.000000pt}%
\pgfpathmoveto{\pgfqpoint{10.084051in}{1.397499in}}%
\pgfpathlineto{\pgfqpoint{9.859813in}{1.432292in}}%
\pgfpathlineto{\pgfqpoint{9.629632in}{1.471279in}}%
\pgfpathlineto{\pgfqpoint{9.412089in}{1.513323in}}%
\pgfpathlineto{\pgfqpoint{9.214079in}{1.558644in}}%
\pgfpathlineto{\pgfqpoint{9.009368in}{1.607450in}}%
\pgfpathlineto{\pgfqpoint{8.800040in}{1.660025in}}%
\pgfpathlineto{\pgfqpoint{8.614649in}{1.716579in}}%
\pgfpathlineto{\pgfqpoint{8.415101in}{1.777362in}}%
\pgfpathlineto{\pgfqpoint{8.227147in}{1.842628in}}%
\pgfpathlineto{\pgfqpoint{8.033843in}{1.912629in}}%
\pgfpathlineto{\pgfqpoint{7.856475in}{1.987635in}}%
\pgfpathlineto{\pgfqpoint{7.683470in}{2.067863in}}%
\pgfpathlineto{\pgfqpoint{7.501072in}{2.153536in}}%
\pgfpathlineto{\pgfqpoint{7.322822in}{2.244848in}}%
\pgfpathlineto{\pgfqpoint{7.154646in}{2.341979in}}%
\pgfpathlineto{\pgfqpoint{6.981560in}{2.445063in}}%
\pgfpathlineto{\pgfqpoint{6.820817in}{2.554175in}}%
\pgfpathlineto{\pgfqpoint{6.651642in}{2.669271in}}%
\pgfpathlineto{\pgfqpoint{6.488591in}{2.790316in}}%
\pgfpathlineto{\pgfqpoint{6.327617in}{2.917139in}}%
\pgfpathlineto{\pgfqpoint{6.167367in}{3.049468in}}%
\pgfpathlineto{\pgfqpoint{6.009181in}{3.186893in}}%
\pgfpathlineto{\pgfqpoint{5.854470in}{3.328876in}}%
\pgfpathlineto{\pgfqpoint{5.706017in}{3.474745in}}%
\pgfpathlineto{\pgfqpoint{5.540124in}{3.623609in}}%
\pgfpathlineto{\pgfqpoint{5.388664in}{3.774465in}}%
\pgfpathlineto{\pgfqpoint{5.231447in}{3.926098in}}%
\pgfpathlineto{\pgfqpoint{5.081052in}{4.077141in}}%
\pgfpathlineto{\pgfqpoint{4.928221in}{4.226088in}}%
\pgfpathlineto{\pgfqpoint{4.775970in}{4.371267in}}%
\pgfpathlineto{\pgfqpoint{4.620484in}{4.510948in}}%
\pgfpathlineto{\pgfqpoint{4.471780in}{4.643250in}}%
\pgfpathlineto{\pgfqpoint{4.320458in}{4.766332in}}%
\pgfpathlineto{\pgfqpoint{4.171013in}{4.878412in}}%
\pgfpathlineto{\pgfqpoint{4.023127in}{4.977764in}}%
\pgfpathlineto{\pgfqpoint{3.878297in}{5.062840in}}%
\pgfpathlineto{\pgfqpoint{3.735969in}{5.132301in}}%
\pgfpathlineto{\pgfqpoint{3.594716in}{5.185098in}}%
\pgfpathlineto{\pgfqpoint{3.454420in}{5.220494in}}%
\pgfpathlineto{\pgfqpoint{3.316192in}{5.238072in}}%
\pgfpathlineto{\pgfqpoint{3.185869in}{5.237845in}}%
\pgfpathlineto{\pgfqpoint{3.054983in}{5.220142in}}%
\pgfpathlineto{\pgfqpoint{2.930120in}{5.185677in}}%
\pgfpathlineto{\pgfqpoint{2.810743in}{5.135475in}}%
\pgfpathlineto{\pgfqpoint{2.698612in}{5.070844in}}%
\pgfpathlineto{\pgfqpoint{2.587676in}{4.993329in}}%
\pgfpathlineto{\pgfqpoint{2.483261in}{4.904644in}}%
\pgfpathlineto{\pgfqpoint{2.385818in}{4.806651in}}%
\pgfpathlineto{\pgfqpoint{2.296163in}{4.701210in}}%
\pgfpathlineto{\pgfqpoint{2.212633in}{4.590281in}}%
\pgfpathlineto{\pgfqpoint{2.133688in}{4.475719in}}%
\pgfpathlineto{\pgfqpoint{2.064329in}{4.359376in}}%
\pgfpathlineto{\pgfqpoint{2.004747in}{4.242978in}}%
\pgfpathlineto{\pgfqpoint{1.949989in}{4.128218in}}%
\pgfpathlineto{\pgfqpoint{1.903752in}{4.016613in}}%
\pgfpathlineto{\pgfqpoint{1.867740in}{3.909584in}}%
\pgfpathlineto{\pgfqpoint{1.839762in}{3.808603in}}%
\pgfpathlineto{\pgfqpoint{1.822914in}{3.714851in}}%
\pgfpathlineto{\pgfqpoint{1.811756in}{3.629578in}}%
\pgfpathlineto{\pgfqpoint{1.813297in}{3.553940in}}%
\pgfpathlineto{\pgfqpoint{1.822694in}{3.489040in}}%
\pgfpathlineto{\pgfqpoint{1.842720in}{3.435760in}}%
\pgfpathlineto{\pgfqpoint{1.872412in}{3.394898in}}%
\pgfpathlineto{\pgfqpoint{1.916245in}{3.367065in}}%
\pgfpathlineto{\pgfqpoint{1.952204in}{3.352237in}}%
\pgfpathlineto{\pgfqpoint{1.999015in}{3.350123in}}%
\pgfpathlineto{\pgfqpoint{2.047110in}{3.359698in}}%
\pgfpathlineto{\pgfqpoint{2.093299in}{3.379191in}}%
\pgfpathlineto{\pgfqpoint{2.134198in}{3.406400in}}%
\pgfpathlineto{\pgfqpoint{2.171639in}{3.438920in}}%
\pgfpathlineto{\pgfqpoint{2.197460in}{3.473965in}}%
\pgfpathlineto{\pgfqpoint{2.219983in}{3.509171in}}%
\pgfpathlineto{\pgfqpoint{2.233192in}{3.542485in}}%
\pgfpathlineto{\pgfqpoint{2.238867in}{3.572415in}}%
\pgfusepath{stroke}%
\end{pgfscope}%
\begin{pgfscope}%
\pgfpathrectangle{\pgfqpoint{0.783051in}{0.693364in}}{\pgfqpoint{9.300000in}{6.795000in}}%
\pgfusepath{clip}%
\pgfsetbuttcap%
\pgfsetroundjoin%
\pgfsetlinewidth{2.007500pt}%
\definecolor{currentstroke}{rgb}{0.160784,0.329412,0.549020}%
\pgfsetstrokecolor{currentstroke}%
\pgfsetdash{{7.400000pt}{3.200000pt}}{0.000000pt}%
\pgfpathmoveto{\pgfqpoint{10.084051in}{1.535487in}}%
\pgfpathlineto{\pgfqpoint{10.078769in}{1.536537in}}%
\pgfpathlineto{\pgfqpoint{9.872458in}{1.586456in}}%
\pgfpathlineto{\pgfqpoint{9.657614in}{1.640555in}}%
\pgfpathlineto{\pgfqpoint{9.467070in}{1.699129in}}%
\pgfpathlineto{\pgfqpoint{9.268857in}{1.762529in}}%
\pgfpathlineto{\pgfqpoint{9.075388in}{1.831116in}}%
\pgfpathlineto{\pgfqpoint{8.888171in}{1.905207in}}%
\pgfpathlineto{\pgfqpoint{8.711484in}{1.985195in}}%
\pgfpathlineto{\pgfqpoint{8.529149in}{2.071445in}}%
\pgfpathlineto{\pgfqpoint{8.349545in}{2.164301in}}%
\pgfpathlineto{\pgfqpoint{8.179676in}{2.264124in}}%
\pgfpathlineto{\pgfqpoint{8.001637in}{2.371252in}}%
\pgfpathlineto{\pgfqpoint{7.837921in}{2.485970in}}%
\pgfpathlineto{\pgfqpoint{7.673707in}{2.608509in}}%
\pgfpathlineto{\pgfqpoint{7.502794in}{2.739102in}}%
\pgfpathlineto{\pgfqpoint{7.339238in}{2.877912in}}%
\pgfpathlineto{\pgfqpoint{7.182588in}{3.024820in}}%
\pgfpathlineto{\pgfqpoint{7.015430in}{3.179809in}}%
\pgfpathlineto{\pgfqpoint{6.857462in}{3.342622in}}%
\pgfpathlineto{\pgfqpoint{6.701115in}{3.512856in}}%
\pgfpathlineto{\pgfqpoint{6.537421in}{3.689922in}}%
\pgfpathlineto{\pgfqpoint{6.381487in}{3.873002in}}%
\pgfpathlineto{\pgfqpoint{6.222679in}{4.061079in}}%
\pgfpathlineto{\pgfqpoint{6.069520in}{4.252863in}}%
\pgfpathlineto{\pgfqpoint{5.903463in}{4.446838in}}%
\pgfpathlineto{\pgfqpoint{5.740905in}{4.641230in}}%
\pgfpathlineto{\pgfqpoint{5.580113in}{4.834021in}}%
\pgfpathlineto{\pgfqpoint{5.419334in}{5.023003in}}%
\pgfpathlineto{\pgfqpoint{5.258621in}{5.205811in}}%
\pgfpathlineto{\pgfqpoint{5.100126in}{5.379944in}}%
\pgfpathlineto{\pgfqpoint{4.936680in}{5.542898in}}%
\pgfpathlineto{\pgfqpoint{4.774221in}{5.692117in}}%
\pgfpathlineto{\pgfqpoint{4.615731in}{5.825262in}}%
\pgfpathlineto{\pgfqpoint{4.454418in}{5.940171in}}%
\pgfpathlineto{\pgfqpoint{4.297814in}{6.035023in}}%
\pgfpathlineto{\pgfqpoint{4.138564in}{6.108303in}}%
\pgfpathlineto{\pgfqpoint{3.984339in}{6.159029in}}%
\pgfpathlineto{\pgfqpoint{3.833582in}{6.186630in}}%
\pgfpathlineto{\pgfqpoint{3.688013in}{6.191126in}}%
\pgfpathlineto{\pgfqpoint{3.539380in}{6.173000in}}%
\pgfpathlineto{\pgfqpoint{3.400594in}{6.133239in}}%
\pgfpathlineto{\pgfqpoint{3.264586in}{6.073251in}}%
\pgfpathlineto{\pgfqpoint{3.135512in}{5.994843in}}%
\pgfpathlineto{\pgfqpoint{3.011689in}{5.900098in}}%
\pgfpathlineto{\pgfqpoint{2.897151in}{5.791348in}}%
\pgfpathlineto{\pgfqpoint{2.785744in}{5.671020in}}%
\pgfpathlineto{\pgfqpoint{2.683641in}{5.541646in}}%
\pgfpathlineto{\pgfqpoint{2.590201in}{5.405725in}}%
\pgfpathlineto{\pgfqpoint{2.503062in}{5.265739in}}%
\pgfpathlineto{\pgfqpoint{2.423819in}{5.124018in}}%
\pgfpathlineto{\pgfqpoint{2.358415in}{4.982819in}}%
\pgfpathlineto{\pgfqpoint{2.296087in}{4.844266in}}%
\pgfpathlineto{\pgfqpoint{2.247327in}{4.710238in}}%
\pgfpathlineto{\pgfqpoint{2.208678in}{4.582708in}}%
\pgfpathlineto{\pgfqpoint{2.177919in}{4.463264in}}%
\pgfpathlineto{\pgfqpoint{2.162049in}{4.353581in}}%
\pgfpathlineto{\pgfqpoint{2.153712in}{4.255071in}}%
\pgfpathlineto{\pgfqpoint{2.158140in}{4.169151in}}%
\pgfpathlineto{\pgfqpoint{2.175679in}{4.097044in}}%
\pgfpathlineto{\pgfqpoint{2.203518in}{4.039806in}}%
\pgfpathlineto{\pgfqpoint{2.234626in}{3.998095in}}%
\pgfpathlineto{\pgfqpoint{2.276649in}{3.972287in}}%
\pgfpathlineto{\pgfqpoint{2.325309in}{3.962025in}}%
\pgfpathlineto{\pgfqpoint{2.378575in}{3.966278in}}%
\pgfpathlineto{\pgfqpoint{2.427330in}{3.983420in}}%
\pgfpathlineto{\pgfqpoint{2.469382in}{4.010926in}}%
\pgfpathlineto{\pgfqpoint{2.512638in}{4.045843in}}%
\pgfpathlineto{\pgfqpoint{2.543995in}{4.085303in}}%
\pgfpathlineto{\pgfqpoint{2.572372in}{4.126136in}}%
\pgfpathlineto{\pgfqpoint{2.588280in}{4.165828in}}%
\pgfpathlineto{\pgfqpoint{2.599825in}{4.202357in}}%
\pgfpathlineto{\pgfqpoint{2.603567in}{4.234299in}}%
\pgfpathlineto{\pgfqpoint{2.601892in}{4.260813in}}%
\pgfusepath{stroke}%
\end{pgfscope}%
\begin{pgfscope}%
\pgfpathrectangle{\pgfqpoint{0.783051in}{0.693364in}}{\pgfqpoint{9.300000in}{6.795000in}}%
\pgfusepath{clip}%
\pgfsetbuttcap%
\pgfsetroundjoin%
\pgfsetlinewidth{2.007500pt}%
\definecolor{currentstroke}{rgb}{0.160784,0.329412,0.549020}%
\pgfsetstrokecolor{currentstroke}%
\pgfsetdash{{7.400000pt}{3.200000pt}}{0.000000pt}%
\pgfpathmoveto{\pgfqpoint{10.084051in}{1.790498in}}%
\pgfpathlineto{\pgfqpoint{10.013843in}{1.814533in}}%
\pgfpathlineto{\pgfqpoint{9.827289in}{1.891438in}}%
\pgfpathlineto{\pgfqpoint{9.659349in}{1.974964in}}%
\pgfpathlineto{\pgfqpoint{9.461241in}{2.065588in}}%
\pgfpathlineto{\pgfqpoint{9.289670in}{2.163814in}}%
\pgfpathlineto{\pgfqpoint{9.108561in}{2.270068in}}%
\pgfpathlineto{\pgfqpoint{8.939076in}{2.384895in}}%
\pgfpathlineto{\pgfqpoint{8.776982in}{2.508734in}}%
\pgfpathlineto{\pgfqpoint{8.600446in}{2.642025in}}%
\pgfpathlineto{\pgfqpoint{8.435836in}{2.785143in}}%
\pgfpathlineto{\pgfqpoint{8.275874in}{2.938416in}}%
\pgfpathlineto{\pgfqpoint{8.113193in}{3.102045in}}%
\pgfpathlineto{\pgfqpoint{7.945734in}{3.276181in}}%
\pgfpathlineto{\pgfqpoint{7.782640in}{3.460821in}}%
\pgfpathlineto{\pgfqpoint{7.624111in}{3.655644in}}%
\pgfpathlineto{\pgfqpoint{7.463980in}{3.860293in}}%
\pgfpathlineto{\pgfqpoint{7.298074in}{4.074096in}}%
\pgfpathlineto{\pgfqpoint{7.135906in}{4.296130in}}%
\pgfpathlineto{\pgfqpoint{6.975115in}{4.525163in}}%
\pgfpathlineto{\pgfqpoint{6.806138in}{4.759651in}}%
\pgfpathlineto{\pgfqpoint{6.641997in}{4.997705in}}%
\pgfpathlineto{\pgfqpoint{6.473754in}{5.237115in}}%
\pgfpathlineto{\pgfqpoint{6.305698in}{5.475364in}}%
\pgfpathlineto{\pgfqpoint{6.148161in}{5.709637in}}%
\pgfpathlineto{\pgfqpoint{5.959809in}{5.936897in}}%
\pgfpathlineto{\pgfqpoint{5.789907in}{6.153956in}}%
\pgfpathlineto{\pgfqpoint{5.610761in}{6.357573in}}%
\pgfpathlineto{\pgfqpoint{5.438912in}{6.544516in}}%
\pgfpathlineto{\pgfqpoint{5.262713in}{6.711781in}}%
\pgfpathlineto{\pgfqpoint{5.084765in}{6.856533in}}%
\pgfpathlineto{\pgfqpoint{4.913713in}{6.976444in}}%
\pgfpathlineto{\pgfqpoint{4.735567in}{7.069643in}}%
\pgfpathlineto{\pgfqpoint{4.567568in}{7.134869in}}%
\pgfpathlineto{\pgfqpoint{4.394475in}{7.171477in}}%
\pgfpathlineto{\pgfqpoint{4.228402in}{7.179500in}}%
\pgfpathlineto{\pgfqpoint{4.067614in}{7.159632in}}%
\pgfpathlineto{\pgfqpoint{3.909090in}{7.113163in}}%
\pgfpathlineto{\pgfqpoint{3.756773in}{7.041969in}}%
\pgfpathlineto{\pgfqpoint{3.611228in}{6.948369in}}%
\pgfpathlineto{\pgfqpoint{3.473837in}{6.835009in}}%
\pgfpathlineto{\pgfqpoint{3.342158in}{6.704852in}}%
\pgfpathlineto{\pgfqpoint{3.217142in}{6.560962in}}%
\pgfpathlineto{\pgfqpoint{3.102594in}{6.406492in}}%
\pgfpathlineto{\pgfqpoint{2.996412in}{6.244544in}}%
\pgfpathlineto{\pgfqpoint{2.899892in}{6.078144in}}%
\pgfpathlineto{\pgfqpoint{2.813110in}{5.910182in}}%
\pgfpathlineto{\pgfqpoint{2.734439in}{5.743384in}}%
\pgfpathlineto{\pgfqpoint{2.669542in}{5.580326in}}%
\pgfpathlineto{\pgfqpoint{2.615469in}{5.423279in}}%
\pgfpathlineto{\pgfqpoint{2.574123in}{5.274592in}}%
\pgfpathlineto{\pgfqpoint{2.545544in}{5.136278in}}%
\pgfpathlineto{\pgfqpoint{2.525875in}{5.010126in}}%
\pgfpathlineto{\pgfqpoint{2.522042in}{4.897987in}}%
\pgfpathlineto{\pgfqpoint{2.528790in}{4.801392in}}%
\pgfpathlineto{\pgfqpoint{2.549181in}{4.721724in}}%
\pgfpathlineto{\pgfqpoint{2.577680in}{4.659948in}}%
\pgfpathlineto{\pgfqpoint{2.618972in}{4.617041in}}%
\pgfpathlineto{\pgfqpoint{2.665925in}{4.592666in}}%
\pgfpathlineto{\pgfqpoint{2.720708in}{4.586291in}}%
\pgfpathlineto{\pgfqpoint{2.773586in}{4.596340in}}%
\pgfpathlineto{\pgfqpoint{2.828748in}{4.620356in}}%
\pgfpathlineto{\pgfqpoint{2.873541in}{4.655295in}}%
\pgfpathlineto{\pgfqpoint{2.914895in}{4.697577in}}%
\pgfpathlineto{\pgfqpoint{2.949891in}{4.743769in}}%
\pgfpathlineto{\pgfqpoint{2.975478in}{4.790498in}}%
\pgfpathlineto{\pgfqpoint{2.991233in}{4.835112in}}%
\pgfpathlineto{\pgfqpoint{3.000824in}{4.875368in}}%
\pgfpathlineto{\pgfqpoint{3.004107in}{4.909972in}}%
\pgfpathlineto{\pgfqpoint{3.005934in}{4.938211in}}%
\pgfpathlineto{\pgfqpoint{2.996841in}{4.959931in}}%
\pgfusepath{stroke}%
\end{pgfscope}%
\begin{pgfscope}%
\pgfsetrectcap%
\pgfsetmiterjoin%
\pgfsetlinewidth{1.003750pt}%
\definecolor{currentstroke}{rgb}{0.000000,0.000000,0.000000}%
\pgfsetstrokecolor{currentstroke}%
\pgfsetdash{}{0pt}%
\pgfpathmoveto{\pgfqpoint{0.783051in}{0.693364in}}%
\pgfpathlineto{\pgfqpoint{0.783051in}{7.488364in}}%
\pgfusepath{stroke}%
\end{pgfscope}%
\begin{pgfscope}%
\pgfsetrectcap%
\pgfsetmiterjoin%
\pgfsetlinewidth{1.003750pt}%
\definecolor{currentstroke}{rgb}{0.000000,0.000000,0.000000}%
\pgfsetstrokecolor{currentstroke}%
\pgfsetdash{}{0pt}%
\pgfpathmoveto{\pgfqpoint{10.083051in}{0.693364in}}%
\pgfpathlineto{\pgfqpoint{10.083051in}{7.488364in}}%
\pgfusepath{stroke}%
\end{pgfscope}%
\begin{pgfscope}%
\pgfsetrectcap%
\pgfsetmiterjoin%
\pgfsetlinewidth{1.003750pt}%
\definecolor{currentstroke}{rgb}{0.000000,0.000000,0.000000}%
\pgfsetstrokecolor{currentstroke}%
\pgfsetdash{}{0pt}%
\pgfpathmoveto{\pgfqpoint{0.783051in}{0.693364in}}%
\pgfpathlineto{\pgfqpoint{10.083051in}{0.693364in}}%
\pgfusepath{stroke}%
\end{pgfscope}%
\begin{pgfscope}%
\pgfsetrectcap%
\pgfsetmiterjoin%
\pgfsetlinewidth{1.003750pt}%
\definecolor{currentstroke}{rgb}{0.000000,0.000000,0.000000}%
\pgfsetstrokecolor{currentstroke}%
\pgfsetdash{}{0pt}%
\pgfpathmoveto{\pgfqpoint{0.783051in}{7.488364in}}%
\pgfpathlineto{\pgfqpoint{10.083051in}{7.488364in}}%
\pgfusepath{stroke}%
\end{pgfscope}%
\begin{pgfscope}%
\definecolor{textcolor}{rgb}{0.372549,0.403922,0.411765}%
\pgfsetstrokecolor{textcolor}%
\pgfsetfillcolor{textcolor}%
\pgftext[x=2.346536in,y=3.085480in,left,base]{\color{textcolor}\rmfamily\fontsize{20.000000}{24.000000}\selectfont \(\displaystyle y = 0 \)}%
\end{pgfscope}%
\begin{pgfscope}%
\definecolor{textcolor}{rgb}{0.372549,0.403922,0.411765}%
\pgfsetstrokecolor{textcolor}%
\pgfsetfillcolor{textcolor}%
\pgftext[x=4.769364in,y=7.180020in,left,base]{\color{textcolor}\rmfamily\fontsize{20.000000}{24.000000}\selectfont \(\displaystyle y = 5 \)}%
\end{pgfscope}%
\begin{pgfscope}%
\pgfsetrectcap%
\pgfsetroundjoin%
\pgfsetlinewidth{2.007500pt}%
\definecolor{currentstroke}{rgb}{0.149020,0.137255,0.133333}%
\pgfsetstrokecolor{currentstroke}%
\pgfsetdash{}{0pt}%
\pgfpathmoveto{\pgfqpoint{7.290596in}{7.175864in}}%
\pgfpathlineto{\pgfqpoint{7.665596in}{7.175864in}}%
\pgfusepath{stroke}%
\end{pgfscope}%
\begin{pgfscope}%
\definecolor{textcolor}{rgb}{0.000000,0.000000,0.000000}%
\pgfsetstrokecolor{textcolor}%
\pgfsetfillcolor{textcolor}%
\pgftext[x=7.865596in,y=7.088364in,left,base]{\color{textcolor}\rmfamily\fontsize{18.000000}{21.600000}\selectfont Free Fermions}%
\end{pgfscope}%
\begin{pgfscope}%
\pgfsetbuttcap%
\pgfsetroundjoin%
\pgfsetlinewidth{2.007500pt}%
\definecolor{currentstroke}{rgb}{0.160784,0.329412,0.549020}%
\pgfsetstrokecolor{currentstroke}%
\pgfsetdash{{7.400000pt}{3.200000pt}}{0.000000pt}%
\pgfpathmoveto{\pgfqpoint{7.290596in}{6.829182in}}%
\pgfpathlineto{\pgfqpoint{7.665596in}{6.829182in}}%
\pgfusepath{stroke}%
\end{pgfscope}%
\begin{pgfscope}%
\definecolor{textcolor}{rgb}{0.000000,0.000000,0.000000}%
\pgfsetstrokecolor{textcolor}%
\pgfsetfillcolor{textcolor}%
\pgftext[x=7.865596in,y=6.741682in,left,base]{\color{textcolor}\rmfamily\fontsize{18.000000}{21.600000}\selectfont Interacting Fermions}%
\end{pgfscope}%
\end{pgfpicture}%
\makeatother%
\endgroup%

%% file: Compac_RG.pgf
\begingroup%
\makeatletter%
\begin{pgfpicture}%
\pgfpathrectangle{\pgfpointorigin}{\pgfqpoint{10.220463in}{7.588364in}}%
\pgfusepath{use as bounding box, clip}%
\begin{pgfscope}%
\pgfsetbuttcap%
\pgfsetmiterjoin%
\definecolor{currentfill}{rgb}{1.000000,1.000000,1.000000}%
\pgfsetfillcolor{currentfill}%
\pgfsetlinewidth{0.000000pt}%
\definecolor{currentstroke}{rgb}{1.000000,1.000000,1.000000}%
\pgfsetstrokecolor{currentstroke}%
\pgfsetdash{}{0pt}%
\pgfpathmoveto{\pgfqpoint{0.000000in}{0.000000in}}%
\pgfpathlineto{\pgfqpoint{10.220463in}{0.000000in}}%
\pgfpathlineto{\pgfqpoint{10.220463in}{7.588364in}}%
\pgfpathlineto{\pgfqpoint{0.000000in}{7.588364in}}%
\pgfpathclose%
\pgfusepath{fill}%
\end{pgfscope}%
\begin{pgfscope}%
\pgfsetbuttcap%
\pgfsetmiterjoin%
\definecolor{currentfill}{rgb}{1.000000,1.000000,1.000000}%
\pgfsetfillcolor{currentfill}%
\pgfsetlinewidth{0.000000pt}%
\definecolor{currentstroke}{rgb}{0.000000,0.000000,0.000000}%
\pgfsetstrokecolor{currentstroke}%
\pgfsetstrokeopacity{0.000000}%
\pgfsetdash{}{0pt}%
\pgfpathmoveto{\pgfqpoint{0.765429in}{0.693364in}}%
\pgfpathlineto{\pgfqpoint{10.065429in}{0.693364in}}%
\pgfpathlineto{\pgfqpoint{10.065429in}{7.488364in}}%
\pgfpathlineto{\pgfqpoint{0.765429in}{7.488364in}}%
\pgfpathclose%
\pgfusepath{fill}%
\end{pgfscope}%
\begin{pgfscope}%
\pgfsetbuttcap%
\pgfsetroundjoin%
\definecolor{currentfill}{rgb}{0.000000,0.000000,0.000000}%
\pgfsetfillcolor{currentfill}%
\pgfsetlinewidth{0.803000pt}%
\definecolor{currentstroke}{rgb}{0.000000,0.000000,0.000000}%
\pgfsetstrokecolor{currentstroke}%
\pgfsetdash{}{0pt}%
\pgfsys@defobject{currentmarker}{\pgfqpoint{0.000000in}{-0.048611in}}{\pgfqpoint{0.000000in}{0.000000in}}{%
\pgfpathmoveto{\pgfqpoint{0.000000in}{0.000000in}}%
\pgfpathlineto{\pgfqpoint{0.000000in}{-0.048611in}}%
\pgfusepath{stroke,fill}%
}%
\begin{pgfscope}%
\pgfsys@transformshift{0.765429in}{0.693364in}%
\pgfsys@useobject{currentmarker}{}%
\end{pgfscope}%
\end{pgfscope}%
\begin{pgfscope}%
\definecolor{textcolor}{rgb}{0.000000,0.000000,0.000000}%
\pgfsetstrokecolor{textcolor}%
\pgfsetfillcolor{textcolor}%
\pgftext[x=0.765429in,y=0.596142in,,top]{\color{textcolor}\rmfamily\fontsize{16.000000}{19.200000}\selectfont \(\displaystyle 0\)}%
\end{pgfscope}%
\begin{pgfscope}%
\pgfsetbuttcap%
\pgfsetroundjoin%
\definecolor{currentfill}{rgb}{0.000000,0.000000,0.000000}%
\pgfsetfillcolor{currentfill}%
\pgfsetlinewidth{0.803000pt}%
\definecolor{currentstroke}{rgb}{0.000000,0.000000,0.000000}%
\pgfsetstrokecolor{currentstroke}%
\pgfsetdash{}{0pt}%
\pgfsys@defobject{currentmarker}{\pgfqpoint{0.000000in}{-0.048611in}}{\pgfqpoint{0.000000in}{0.000000in}}{%
\pgfpathmoveto{\pgfqpoint{0.000000in}{0.000000in}}%
\pgfpathlineto{\pgfqpoint{0.000000in}{-0.048611in}}%
\pgfusepath{stroke,fill}%
}%
\begin{pgfscope}%
\pgfsys@transformshift{1.798762in}{0.693364in}%
\pgfsys@useobject{currentmarker}{}%
\end{pgfscope}%
\end{pgfscope}%
\begin{pgfscope}%
\definecolor{textcolor}{rgb}{0.000000,0.000000,0.000000}%
\pgfsetstrokecolor{textcolor}%
\pgfsetfillcolor{textcolor}%
\pgftext[x=1.798762in,y=0.596142in,,top]{\color{textcolor}\rmfamily\fontsize{16.000000}{19.200000}\selectfont \(\displaystyle 1\)}%
\end{pgfscope}%
\begin{pgfscope}%
\pgfsetbuttcap%
\pgfsetroundjoin%
\definecolor{currentfill}{rgb}{0.000000,0.000000,0.000000}%
\pgfsetfillcolor{currentfill}%
\pgfsetlinewidth{0.803000pt}%
\definecolor{currentstroke}{rgb}{0.000000,0.000000,0.000000}%
\pgfsetstrokecolor{currentstroke}%
\pgfsetdash{}{0pt}%
\pgfsys@defobject{currentmarker}{\pgfqpoint{0.000000in}{-0.048611in}}{\pgfqpoint{0.000000in}{0.000000in}}{%
\pgfpathmoveto{\pgfqpoint{0.000000in}{0.000000in}}%
\pgfpathlineto{\pgfqpoint{0.000000in}{-0.048611in}}%
\pgfusepath{stroke,fill}%
}%
\begin{pgfscope}%
\pgfsys@transformshift{2.832095in}{0.693364in}%
\pgfsys@useobject{currentmarker}{}%
\end{pgfscope}%
\end{pgfscope}%
\begin{pgfscope}%
\definecolor{textcolor}{rgb}{0.000000,0.000000,0.000000}%
\pgfsetstrokecolor{textcolor}%
\pgfsetfillcolor{textcolor}%
\pgftext[x=2.832095in,y=0.596142in,,top]{\color{textcolor}\rmfamily\fontsize{16.000000}{19.200000}\selectfont \(\displaystyle 2\)}%
\end{pgfscope}%
\begin{pgfscope}%
\pgfsetbuttcap%
\pgfsetroundjoin%
\definecolor{currentfill}{rgb}{0.000000,0.000000,0.000000}%
\pgfsetfillcolor{currentfill}%
\pgfsetlinewidth{0.803000pt}%
\definecolor{currentstroke}{rgb}{0.000000,0.000000,0.000000}%
\pgfsetstrokecolor{currentstroke}%
\pgfsetdash{}{0pt}%
\pgfsys@defobject{currentmarker}{\pgfqpoint{0.000000in}{-0.048611in}}{\pgfqpoint{0.000000in}{0.000000in}}{%
\pgfpathmoveto{\pgfqpoint{0.000000in}{0.000000in}}%
\pgfpathlineto{\pgfqpoint{0.000000in}{-0.048611in}}%
\pgfusepath{stroke,fill}%
}%
\begin{pgfscope}%
\pgfsys@transformshift{3.865429in}{0.693364in}%
\pgfsys@useobject{currentmarker}{}%
\end{pgfscope}%
\end{pgfscope}%
\begin{pgfscope}%
\definecolor{textcolor}{rgb}{0.000000,0.000000,0.000000}%
\pgfsetstrokecolor{textcolor}%
\pgfsetfillcolor{textcolor}%
\pgftext[x=3.865429in,y=0.596142in,,top]{\color{textcolor}\rmfamily\fontsize{16.000000}{19.200000}\selectfont \(\displaystyle 3\)}%
\end{pgfscope}%
\begin{pgfscope}%
\pgfsetbuttcap%
\pgfsetroundjoin%
\definecolor{currentfill}{rgb}{0.000000,0.000000,0.000000}%
\pgfsetfillcolor{currentfill}%
\pgfsetlinewidth{0.803000pt}%
\definecolor{currentstroke}{rgb}{0.000000,0.000000,0.000000}%
\pgfsetstrokecolor{currentstroke}%
\pgfsetdash{}{0pt}%
\pgfsys@defobject{currentmarker}{\pgfqpoint{0.000000in}{-0.048611in}}{\pgfqpoint{0.000000in}{0.000000in}}{%
\pgfpathmoveto{\pgfqpoint{0.000000in}{0.000000in}}%
\pgfpathlineto{\pgfqpoint{0.000000in}{-0.048611in}}%
\pgfusepath{stroke,fill}%
}%
\begin{pgfscope}%
\pgfsys@transformshift{4.898762in}{0.693364in}%
\pgfsys@useobject{currentmarker}{}%
\end{pgfscope}%
\end{pgfscope}%
\begin{pgfscope}%
\definecolor{textcolor}{rgb}{0.000000,0.000000,0.000000}%
\pgfsetstrokecolor{textcolor}%
\pgfsetfillcolor{textcolor}%
\pgftext[x=4.898762in,y=0.596142in,,top]{\color{textcolor}\rmfamily\fontsize{16.000000}{19.200000}\selectfont \(\displaystyle 4\)}%
\end{pgfscope}%
\begin{pgfscope}%
\pgfsetbuttcap%
\pgfsetroundjoin%
\definecolor{currentfill}{rgb}{0.000000,0.000000,0.000000}%
\pgfsetfillcolor{currentfill}%
\pgfsetlinewidth{0.803000pt}%
\definecolor{currentstroke}{rgb}{0.000000,0.000000,0.000000}%
\pgfsetstrokecolor{currentstroke}%
\pgfsetdash{}{0pt}%
\pgfsys@defobject{currentmarker}{\pgfqpoint{0.000000in}{-0.048611in}}{\pgfqpoint{0.000000in}{0.000000in}}{%
\pgfpathmoveto{\pgfqpoint{0.000000in}{0.000000in}}%
\pgfpathlineto{\pgfqpoint{0.000000in}{-0.048611in}}%
\pgfusepath{stroke,fill}%
}%
\begin{pgfscope}%
\pgfsys@transformshift{5.932095in}{0.693364in}%
\pgfsys@useobject{currentmarker}{}%
\end{pgfscope}%
\end{pgfscope}%
\begin{pgfscope}%
\definecolor{textcolor}{rgb}{0.000000,0.000000,0.000000}%
\pgfsetstrokecolor{textcolor}%
\pgfsetfillcolor{textcolor}%
\pgftext[x=5.932095in,y=0.596142in,,top]{\color{textcolor}\rmfamily\fontsize{16.000000}{19.200000}\selectfont \(\displaystyle 5\)}%
\end{pgfscope}%
\begin{pgfscope}%
\pgfsetbuttcap%
\pgfsetroundjoin%
\definecolor{currentfill}{rgb}{0.000000,0.000000,0.000000}%
\pgfsetfillcolor{currentfill}%
\pgfsetlinewidth{0.803000pt}%
\definecolor{currentstroke}{rgb}{0.000000,0.000000,0.000000}%
\pgfsetstrokecolor{currentstroke}%
\pgfsetdash{}{0pt}%
\pgfsys@defobject{currentmarker}{\pgfqpoint{0.000000in}{-0.048611in}}{\pgfqpoint{0.000000in}{0.000000in}}{%
\pgfpathmoveto{\pgfqpoint{0.000000in}{0.000000in}}%
\pgfpathlineto{\pgfqpoint{0.000000in}{-0.048611in}}%
\pgfusepath{stroke,fill}%
}%
\begin{pgfscope}%
\pgfsys@transformshift{6.965429in}{0.693364in}%
\pgfsys@useobject{currentmarker}{}%
\end{pgfscope}%
\end{pgfscope}%
\begin{pgfscope}%
\definecolor{textcolor}{rgb}{0.000000,0.000000,0.000000}%
\pgfsetstrokecolor{textcolor}%
\pgfsetfillcolor{textcolor}%
\pgftext[x=6.965429in,y=0.596142in,,top]{\color{textcolor}\rmfamily\fontsize{16.000000}{19.200000}\selectfont \(\displaystyle 6\)}%
\end{pgfscope}%
\begin{pgfscope}%
\pgfsetbuttcap%
\pgfsetroundjoin%
\definecolor{currentfill}{rgb}{0.000000,0.000000,0.000000}%
\pgfsetfillcolor{currentfill}%
\pgfsetlinewidth{0.803000pt}%
\definecolor{currentstroke}{rgb}{0.000000,0.000000,0.000000}%
\pgfsetstrokecolor{currentstroke}%
\pgfsetdash{}{0pt}%
\pgfsys@defobject{currentmarker}{\pgfqpoint{0.000000in}{-0.048611in}}{\pgfqpoint{0.000000in}{0.000000in}}{%
\pgfpathmoveto{\pgfqpoint{0.000000in}{0.000000in}}%
\pgfpathlineto{\pgfqpoint{0.000000in}{-0.048611in}}%
\pgfusepath{stroke,fill}%
}%
\begin{pgfscope}%
\pgfsys@transformshift{7.998762in}{0.693364in}%
\pgfsys@useobject{currentmarker}{}%
\end{pgfscope}%
\end{pgfscope}%
\begin{pgfscope}%
\definecolor{textcolor}{rgb}{0.000000,0.000000,0.000000}%
\pgfsetstrokecolor{textcolor}%
\pgfsetfillcolor{textcolor}%
\pgftext[x=7.998762in,y=0.596142in,,top]{\color{textcolor}\rmfamily\fontsize{16.000000}{19.200000}\selectfont \(\displaystyle 7\)}%
\end{pgfscope}%
\begin{pgfscope}%
\pgfsetbuttcap%
\pgfsetroundjoin%
\definecolor{currentfill}{rgb}{0.000000,0.000000,0.000000}%
\pgfsetfillcolor{currentfill}%
\pgfsetlinewidth{0.803000pt}%
\definecolor{currentstroke}{rgb}{0.000000,0.000000,0.000000}%
\pgfsetstrokecolor{currentstroke}%
\pgfsetdash{}{0pt}%
\pgfsys@defobject{currentmarker}{\pgfqpoint{0.000000in}{-0.048611in}}{\pgfqpoint{0.000000in}{0.000000in}}{%
\pgfpathmoveto{\pgfqpoint{0.000000in}{0.000000in}}%
\pgfpathlineto{\pgfqpoint{0.000000in}{-0.048611in}}%
\pgfusepath{stroke,fill}%
}%
\begin{pgfscope}%
\pgfsys@transformshift{9.032095in}{0.693364in}%
\pgfsys@useobject{currentmarker}{}%
\end{pgfscope}%
\end{pgfscope}%
\begin{pgfscope}%
\definecolor{textcolor}{rgb}{0.000000,0.000000,0.000000}%
\pgfsetstrokecolor{textcolor}%
\pgfsetfillcolor{textcolor}%
\pgftext[x=9.032095in,y=0.596142in,,top]{\color{textcolor}\rmfamily\fontsize{16.000000}{19.200000}\selectfont \(\displaystyle 8\)}%
\end{pgfscope}%
\begin{pgfscope}%
\pgfsetbuttcap%
\pgfsetroundjoin%
\definecolor{currentfill}{rgb}{0.000000,0.000000,0.000000}%
\pgfsetfillcolor{currentfill}%
\pgfsetlinewidth{0.803000pt}%
\definecolor{currentstroke}{rgb}{0.000000,0.000000,0.000000}%
\pgfsetstrokecolor{currentstroke}%
\pgfsetdash{}{0pt}%
\pgfsys@defobject{currentmarker}{\pgfqpoint{0.000000in}{-0.048611in}}{\pgfqpoint{0.000000in}{0.000000in}}{%
\pgfpathmoveto{\pgfqpoint{0.000000in}{0.000000in}}%
\pgfpathlineto{\pgfqpoint{0.000000in}{-0.048611in}}%
\pgfusepath{stroke,fill}%
}%
\begin{pgfscope}%
\pgfsys@transformshift{10.065429in}{0.693364in}%
\pgfsys@useobject{currentmarker}{}%
\end{pgfscope}%
\end{pgfscope}%
\begin{pgfscope}%
\definecolor{textcolor}{rgb}{0.000000,0.000000,0.000000}%
\pgfsetstrokecolor{textcolor}%
\pgfsetfillcolor{textcolor}%
\pgftext[x=10.065429in,y=0.596142in,,top]{\color{textcolor}\rmfamily\fontsize{16.000000}{19.200000}\selectfont \(\displaystyle 9\)}%
\end{pgfscope}%
\begin{pgfscope}%
\definecolor{textcolor}{rgb}{0.000000,0.000000,0.000000}%
\pgfsetstrokecolor{textcolor}%
\pgfsetfillcolor{textcolor}%
\pgftext[x=5.415429in,y=0.313349in,,top]{\color{textcolor}\rmfamily\fontsize{18.000000}{21.600000}\selectfont \(\displaystyle y\)}%
\end{pgfscope}%
\begin{pgfscope}%
\pgfsetbuttcap%
\pgfsetroundjoin%
\definecolor{currentfill}{rgb}{0.000000,0.000000,0.000000}%
\pgfsetfillcolor{currentfill}%
\pgfsetlinewidth{0.803000pt}%
\definecolor{currentstroke}{rgb}{0.000000,0.000000,0.000000}%
\pgfsetstrokecolor{currentstroke}%
\pgfsetdash{}{0pt}%
\pgfsys@defobject{currentmarker}{\pgfqpoint{-0.048611in}{0.000000in}}{\pgfqpoint{0.000000in}{0.000000in}}{%
\pgfpathmoveto{\pgfqpoint{0.000000in}{0.000000in}}%
\pgfpathlineto{\pgfqpoint{-0.048611in}{0.000000in}}%
\pgfusepath{stroke,fill}%
}%
\begin{pgfscope}%
\pgfsys@transformshift{0.765429in}{0.693364in}%
\pgfsys@useobject{currentmarker}{}%
\end{pgfscope}%
\end{pgfscope}%
\begin{pgfscope}%
\definecolor{textcolor}{rgb}{0.000000,0.000000,0.000000}%
\pgfsetstrokecolor{textcolor}%
\pgfsetfillcolor{textcolor}%
\pgftext[x=0.382793in,y=0.610031in,left,base]{\color{textcolor}\rmfamily\fontsize{16.000000}{19.200000}\selectfont \(\displaystyle 0.0\)}%
\end{pgfscope}%
\begin{pgfscope}%
\pgfsetbuttcap%
\pgfsetroundjoin%
\definecolor{currentfill}{rgb}{0.000000,0.000000,0.000000}%
\pgfsetfillcolor{currentfill}%
\pgfsetlinewidth{0.803000pt}%
\definecolor{currentstroke}{rgb}{0.000000,0.000000,0.000000}%
\pgfsetstrokecolor{currentstroke}%
\pgfsetdash{}{0pt}%
\pgfsys@defobject{currentmarker}{\pgfqpoint{-0.048611in}{0.000000in}}{\pgfqpoint{0.000000in}{0.000000in}}{%
\pgfpathmoveto{\pgfqpoint{0.000000in}{0.000000in}}%
\pgfpathlineto{\pgfqpoint{-0.048611in}{0.000000in}}%
\pgfusepath{stroke,fill}%
}%
\begin{pgfscope}%
\pgfsys@transformshift{0.765429in}{1.928818in}%
\pgfsys@useobject{currentmarker}{}%
\end{pgfscope}%
\end{pgfscope}%
\begin{pgfscope}%
\definecolor{textcolor}{rgb}{0.000000,0.000000,0.000000}%
\pgfsetstrokecolor{textcolor}%
\pgfsetfillcolor{textcolor}%
\pgftext[x=0.382793in,y=1.845485in,left,base]{\color{textcolor}\rmfamily\fontsize{16.000000}{19.200000}\selectfont \(\displaystyle 0.1\)}%
\end{pgfscope}%
\begin{pgfscope}%
\pgfsetbuttcap%
\pgfsetroundjoin%
\definecolor{currentfill}{rgb}{0.000000,0.000000,0.000000}%
\pgfsetfillcolor{currentfill}%
\pgfsetlinewidth{0.803000pt}%
\definecolor{currentstroke}{rgb}{0.000000,0.000000,0.000000}%
\pgfsetstrokecolor{currentstroke}%
\pgfsetdash{}{0pt}%
\pgfsys@defobject{currentmarker}{\pgfqpoint{-0.048611in}{0.000000in}}{\pgfqpoint{0.000000in}{0.000000in}}{%
\pgfpathmoveto{\pgfqpoint{0.000000in}{0.000000in}}%
\pgfpathlineto{\pgfqpoint{-0.048611in}{0.000000in}}%
\pgfusepath{stroke,fill}%
}%
\begin{pgfscope}%
\pgfsys@transformshift{0.765429in}{3.164273in}%
\pgfsys@useobject{currentmarker}{}%
\end{pgfscope}%
\end{pgfscope}%
\begin{pgfscope}%
\definecolor{textcolor}{rgb}{0.000000,0.000000,0.000000}%
\pgfsetstrokecolor{textcolor}%
\pgfsetfillcolor{textcolor}%
\pgftext[x=0.382793in,y=3.080940in,left,base]{\color{textcolor}\rmfamily\fontsize{16.000000}{19.200000}\selectfont \(\displaystyle 0.2\)}%
\end{pgfscope}%
\begin{pgfscope}%
\pgfsetbuttcap%
\pgfsetroundjoin%
\definecolor{currentfill}{rgb}{0.000000,0.000000,0.000000}%
\pgfsetfillcolor{currentfill}%
\pgfsetlinewidth{0.803000pt}%
\definecolor{currentstroke}{rgb}{0.000000,0.000000,0.000000}%
\pgfsetstrokecolor{currentstroke}%
\pgfsetdash{}{0pt}%
\pgfsys@defobject{currentmarker}{\pgfqpoint{-0.048611in}{0.000000in}}{\pgfqpoint{0.000000in}{0.000000in}}{%
\pgfpathmoveto{\pgfqpoint{0.000000in}{0.000000in}}%
\pgfpathlineto{\pgfqpoint{-0.048611in}{0.000000in}}%
\pgfusepath{stroke,fill}%
}%
\begin{pgfscope}%
\pgfsys@transformshift{0.765429in}{4.399728in}%
\pgfsys@useobject{currentmarker}{}%
\end{pgfscope}%
\end{pgfscope}%
\begin{pgfscope}%
\definecolor{textcolor}{rgb}{0.000000,0.000000,0.000000}%
\pgfsetstrokecolor{textcolor}%
\pgfsetfillcolor{textcolor}%
\pgftext[x=0.382793in,y=4.316394in,left,base]{\color{textcolor}\rmfamily\fontsize{16.000000}{19.200000}\selectfont \(\displaystyle 0.3\)}%
\end{pgfscope}%
\begin{pgfscope}%
\pgfsetbuttcap%
\pgfsetroundjoin%
\definecolor{currentfill}{rgb}{0.000000,0.000000,0.000000}%
\pgfsetfillcolor{currentfill}%
\pgfsetlinewidth{0.803000pt}%
\definecolor{currentstroke}{rgb}{0.000000,0.000000,0.000000}%
\pgfsetstrokecolor{currentstroke}%
\pgfsetdash{}{0pt}%
\pgfsys@defobject{currentmarker}{\pgfqpoint{-0.048611in}{0.000000in}}{\pgfqpoint{0.000000in}{0.000000in}}{%
\pgfpathmoveto{\pgfqpoint{0.000000in}{0.000000in}}%
\pgfpathlineto{\pgfqpoint{-0.048611in}{0.000000in}}%
\pgfusepath{stroke,fill}%
}%
\begin{pgfscope}%
\pgfsys@transformshift{0.765429in}{5.635182in}%
\pgfsys@useobject{currentmarker}{}%
\end{pgfscope}%
\end{pgfscope}%
\begin{pgfscope}%
\definecolor{textcolor}{rgb}{0.000000,0.000000,0.000000}%
\pgfsetstrokecolor{textcolor}%
\pgfsetfillcolor{textcolor}%
\pgftext[x=0.382793in,y=5.551849in,left,base]{\color{textcolor}\rmfamily\fontsize{16.000000}{19.200000}\selectfont \(\displaystyle 0.4\)}%
\end{pgfscope}%
\begin{pgfscope}%
\pgfsetbuttcap%
\pgfsetroundjoin%
\definecolor{currentfill}{rgb}{0.000000,0.000000,0.000000}%
\pgfsetfillcolor{currentfill}%
\pgfsetlinewidth{0.803000pt}%
\definecolor{currentstroke}{rgb}{0.000000,0.000000,0.000000}%
\pgfsetstrokecolor{currentstroke}%
\pgfsetdash{}{0pt}%
\pgfsys@defobject{currentmarker}{\pgfqpoint{-0.048611in}{0.000000in}}{\pgfqpoint{0.000000in}{0.000000in}}{%
\pgfpathmoveto{\pgfqpoint{0.000000in}{0.000000in}}%
\pgfpathlineto{\pgfqpoint{-0.048611in}{0.000000in}}%
\pgfusepath{stroke,fill}%
}%
\begin{pgfscope}%
\pgfsys@transformshift{0.765429in}{6.870637in}%
\pgfsys@useobject{currentmarker}{}%
\end{pgfscope}%
\end{pgfscope}%
\begin{pgfscope}%
\definecolor{textcolor}{rgb}{0.000000,0.000000,0.000000}%
\pgfsetstrokecolor{textcolor}%
\pgfsetfillcolor{textcolor}%
\pgftext[x=0.382793in,y=6.787303in,left,base]{\color{textcolor}\rmfamily\fontsize{16.000000}{19.200000}\selectfont \(\displaystyle 0.5\)}%
\end{pgfscope}%
\begin{pgfscope}%
\definecolor{textcolor}{rgb}{0.000000,0.000000,0.000000}%
\pgfsetstrokecolor{textcolor}%
\pgfsetfillcolor{textcolor}%
\pgftext[x=0.313349in,y=4.090864in,,bottom,rotate=90.000000]{\color{textcolor}\rmfamily\fontsize{18.000000}{21.600000}\selectfont \(\displaystyle C\)}%
\end{pgfscope}%
\begin{pgfscope}%
\pgfpathrectangle{\pgfqpoint{0.765429in}{0.693364in}}{\pgfqpoint{9.300000in}{6.795000in}}%
\pgfusepath{clip}%
\pgfsetrectcap%
\pgfsetroundjoin%
\pgfsetlinewidth{2.007500pt}%
\definecolor{currentstroke}{rgb}{0.149020,0.137255,0.133333}%
\pgfsetstrokecolor{currentstroke}%
\pgfsetdash{}{0pt}%
\pgfpathmoveto{\pgfqpoint{0.765429in}{2.107395in}}%
\pgfpathlineto{\pgfqpoint{0.806762in}{2.111297in}}%
\pgfpathlineto{\pgfqpoint{0.848095in}{2.115408in}}%
\pgfpathlineto{\pgfqpoint{0.889429in}{2.119727in}}%
\pgfpathlineto{\pgfqpoint{0.930762in}{2.124254in}}%
\pgfpathlineto{\pgfqpoint{0.972095in}{2.128990in}}%
\pgfpathlineto{\pgfqpoint{1.013429in}{2.133934in}}%
\pgfpathlineto{\pgfqpoint{1.054762in}{2.139087in}}%
\pgfpathlineto{\pgfqpoint{1.096095in}{2.144448in}}%
\pgfpathlineto{\pgfqpoint{1.137429in}{2.150018in}}%
\pgfpathlineto{\pgfqpoint{1.178762in}{2.155796in}}%
\pgfpathlineto{\pgfqpoint{1.220095in}{2.161782in}}%
\pgfpathlineto{\pgfqpoint{1.261429in}{2.167977in}}%
\pgfpathlineto{\pgfqpoint{1.302762in}{2.174380in}}%
\pgfpathlineto{\pgfqpoint{1.344095in}{2.180991in}}%
\pgfpathlineto{\pgfqpoint{1.385429in}{2.187811in}}%
\pgfpathlineto{\pgfqpoint{1.426762in}{2.194840in}}%
\pgfpathlineto{\pgfqpoint{1.468095in}{2.202076in}}%
\pgfpathlineto{\pgfqpoint{1.509429in}{2.209521in}}%
\pgfpathlineto{\pgfqpoint{1.550762in}{2.217175in}}%
\pgfpathlineto{\pgfqpoint{1.592095in}{2.225037in}}%
\pgfpathlineto{\pgfqpoint{1.633429in}{2.233107in}}%
\pgfpathlineto{\pgfqpoint{1.674762in}{2.241386in}}%
\pgfpathlineto{\pgfqpoint{1.716095in}{2.249873in}}%
\pgfpathlineto{\pgfqpoint{1.757429in}{2.258569in}}%
\pgfpathlineto{\pgfqpoint{1.798762in}{2.267472in}}%
\pgfpathlineto{\pgfqpoint{1.860762in}{2.286767in}}%
\pgfpathlineto{\pgfqpoint{1.922762in}{2.305822in}}%
\pgfpathlineto{\pgfqpoint{1.984762in}{2.324637in}}%
\pgfpathlineto{\pgfqpoint{2.046762in}{2.343213in}}%
\pgfpathlineto{\pgfqpoint{2.108762in}{2.361550in}}%
\pgfpathlineto{\pgfqpoint{2.170762in}{2.379648in}}%
\pgfpathlineto{\pgfqpoint{2.232762in}{2.397506in}}%
\pgfpathlineto{\pgfqpoint{2.294762in}{2.415125in}}%
\pgfpathlineto{\pgfqpoint{2.356762in}{2.432505in}}%
\pgfpathlineto{\pgfqpoint{2.418762in}{2.449646in}}%
\pgfpathlineto{\pgfqpoint{2.480762in}{2.466547in}}%
\pgfpathlineto{\pgfqpoint{2.542762in}{2.483209in}}%
\pgfpathlineto{\pgfqpoint{2.604762in}{2.499631in}}%
\pgfpathlineto{\pgfqpoint{2.666762in}{2.515814in}}%
\pgfpathlineto{\pgfqpoint{2.728762in}{2.531758in}}%
\pgfpathlineto{\pgfqpoint{2.790762in}{2.547463in}}%
\pgfpathlineto{\pgfqpoint{2.863095in}{2.565723in}}%
\pgfpathlineto{\pgfqpoint{2.914762in}{2.578761in}}%
\pgfpathlineto{\pgfqpoint{2.966429in}{2.591593in}}%
\pgfpathlineto{\pgfqpoint{3.018095in}{2.604217in}}%
\pgfpathlineto{\pgfqpoint{3.069762in}{2.616633in}}%
\pgfpathlineto{\pgfqpoint{3.121429in}{2.628843in}}%
\pgfpathlineto{\pgfqpoint{3.173095in}{2.640844in}}%
\pgfpathlineto{\pgfqpoint{3.224762in}{2.652638in}}%
\pgfpathlineto{\pgfqpoint{3.276429in}{2.664225in}}%
\pgfpathlineto{\pgfqpoint{3.328095in}{2.675605in}}%
\pgfpathlineto{\pgfqpoint{3.379762in}{2.686777in}}%
\pgfpathlineto{\pgfqpoint{3.431429in}{2.697741in}}%
\pgfpathlineto{\pgfqpoint{3.483095in}{2.708498in}}%
\pgfpathlineto{\pgfqpoint{3.534762in}{2.719048in}}%
\pgfpathlineto{\pgfqpoint{3.586429in}{2.729390in}}%
\pgfpathlineto{\pgfqpoint{3.638095in}{2.739525in}}%
\pgfpathlineto{\pgfqpoint{3.689762in}{2.749452in}}%
\pgfpathlineto{\pgfqpoint{3.741429in}{2.759172in}}%
\pgfpathlineto{\pgfqpoint{3.793095in}{2.768684in}}%
\pgfpathlineto{\pgfqpoint{3.844762in}{2.777989in}}%
\pgfpathlineto{\pgfqpoint{3.865429in}{2.781653in}}%
\pgfpathlineto{\pgfqpoint{3.937762in}{2.793090in}}%
\pgfpathlineto{\pgfqpoint{4.010095in}{2.804290in}}%
\pgfpathlineto{\pgfqpoint{4.082429in}{2.815252in}}%
\pgfpathlineto{\pgfqpoint{4.154762in}{2.825978in}}%
\pgfpathlineto{\pgfqpoint{4.227095in}{2.836466in}}%
\pgfpathlineto{\pgfqpoint{4.299429in}{2.846717in}}%
\pgfpathlineto{\pgfqpoint{4.371762in}{2.856731in}}%
\pgfpathlineto{\pgfqpoint{4.444095in}{2.866508in}}%
\pgfpathlineto{\pgfqpoint{4.516429in}{2.876047in}}%
\pgfpathlineto{\pgfqpoint{4.588762in}{2.885350in}}%
\pgfpathlineto{\pgfqpoint{4.661095in}{2.894415in}}%
\pgfpathlineto{\pgfqpoint{4.733429in}{2.903243in}}%
\pgfpathlineto{\pgfqpoint{4.805762in}{2.911834in}}%
\pgfpathlineto{\pgfqpoint{4.878095in}{2.920188in}}%
\pgfpathlineto{\pgfqpoint{4.909095in}{2.923603in}}%
\pgfpathlineto{\pgfqpoint{5.002095in}{2.933114in}}%
\pgfpathlineto{\pgfqpoint{5.095095in}{2.942385in}}%
\pgfpathlineto{\pgfqpoint{5.188095in}{2.951416in}}%
\pgfpathlineto{\pgfqpoint{5.281095in}{2.960207in}}%
\pgfpathlineto{\pgfqpoint{5.374095in}{2.968757in}}%
\pgfpathlineto{\pgfqpoint{5.467095in}{2.977066in}}%
\pgfpathlineto{\pgfqpoint{5.560095in}{2.985136in}}%
\pgfpathlineto{\pgfqpoint{5.653095in}{2.992965in}}%
\pgfpathlineto{\pgfqpoint{5.746095in}{3.000553in}}%
\pgfpathlineto{\pgfqpoint{5.839095in}{3.007902in}}%
\pgfpathlineto{\pgfqpoint{5.932095in}{3.015010in}}%
\pgfpathlineto{\pgfqpoint{6.045762in}{3.022752in}}%
\pgfpathlineto{\pgfqpoint{6.159429in}{3.030287in}}%
\pgfpathlineto{\pgfqpoint{6.273095in}{3.037617in}}%
\pgfpathlineto{\pgfqpoint{6.386762in}{3.044741in}}%
\pgfpathlineto{\pgfqpoint{6.500429in}{3.051659in}}%
\pgfpathlineto{\pgfqpoint{6.614095in}{3.058372in}}%
\pgfpathlineto{\pgfqpoint{6.727762in}{3.064878in}}%
\pgfpathlineto{\pgfqpoint{6.841429in}{3.071178in}}%
\pgfpathlineto{\pgfqpoint{6.955095in}{3.077273in}}%
\pgfpathlineto{\pgfqpoint{7.006762in}{3.079902in}}%
\pgfpathlineto{\pgfqpoint{7.141095in}{3.086534in}}%
\pgfpathlineto{\pgfqpoint{7.275429in}{3.092942in}}%
\pgfpathlineto{\pgfqpoint{7.409762in}{3.099127in}}%
\pgfpathlineto{\pgfqpoint{7.544095in}{3.105088in}}%
\pgfpathlineto{\pgfqpoint{7.678429in}{3.110826in}}%
\pgfpathlineto{\pgfqpoint{7.812762in}{3.116340in}}%
\pgfpathlineto{\pgfqpoint{7.947095in}{3.121631in}}%
\pgfpathlineto{\pgfqpoint{8.019429in}{3.124337in}}%
\pgfpathlineto{\pgfqpoint{8.195095in}{3.130417in}}%
\pgfpathlineto{\pgfqpoint{8.370762in}{3.136263in}}%
\pgfpathlineto{\pgfqpoint{8.546429in}{3.141875in}}%
\pgfpathlineto{\pgfqpoint{8.722095in}{3.147253in}}%
\pgfpathlineto{\pgfqpoint{8.897762in}{3.152396in}}%
\pgfpathlineto{\pgfqpoint{9.042429in}{3.156437in}}%
\pgfpathlineto{\pgfqpoint{9.280095in}{3.162430in}}%
\pgfpathlineto{\pgfqpoint{9.517762in}{3.168199in}}%
\pgfpathlineto{\pgfqpoint{9.755429in}{3.173744in}}%
\pgfpathlineto{\pgfqpoint{9.993095in}{3.179065in}}%
\pgfpathlineto{\pgfqpoint{10.066429in}{3.180661in}}%
\pgfpathlineto{\pgfqpoint{10.066429in}{3.180661in}}%
\pgfusepath{stroke}%
\end{pgfscope}%
\begin{pgfscope}%
\pgfpathrectangle{\pgfqpoint{0.765429in}{0.693364in}}{\pgfqpoint{9.300000in}{6.795000in}}%
\pgfusepath{clip}%
\pgfsetbuttcap%
\pgfsetroundjoin%
\pgfsetlinewidth{2.007500pt}%
\definecolor{currentstroke}{rgb}{0.858824,0.721569,0.368627}%
\pgfsetstrokecolor{currentstroke}%
\pgfsetdash{{16.000000pt}{6.000000pt}}{0.000000pt}%
\pgfpathmoveto{\pgfqpoint{0.765429in}{1.984508in}}%
\pgfpathlineto{\pgfqpoint{0.806762in}{1.987819in}}%
\pgfpathlineto{\pgfqpoint{0.848095in}{1.991410in}}%
\pgfpathlineto{\pgfqpoint{0.889429in}{1.995279in}}%
\pgfpathlineto{\pgfqpoint{0.930762in}{1.999428in}}%
\pgfpathlineto{\pgfqpoint{0.972095in}{2.003855in}}%
\pgfpathlineto{\pgfqpoint{1.013429in}{2.008560in}}%
\pgfpathlineto{\pgfqpoint{1.054762in}{2.013545in}}%
\pgfpathlineto{\pgfqpoint{1.096095in}{2.018808in}}%
\pgfpathlineto{\pgfqpoint{1.137429in}{2.024350in}}%
\pgfpathlineto{\pgfqpoint{1.178762in}{2.030171in}}%
\pgfpathlineto{\pgfqpoint{1.220095in}{2.036271in}}%
\pgfpathlineto{\pgfqpoint{1.261429in}{2.042650in}}%
\pgfpathlineto{\pgfqpoint{1.302762in}{2.049307in}}%
\pgfpathlineto{\pgfqpoint{1.344095in}{2.056243in}}%
\pgfpathlineto{\pgfqpoint{1.385429in}{2.063458in}}%
\pgfpathlineto{\pgfqpoint{1.426762in}{2.070952in}}%
\pgfpathlineto{\pgfqpoint{1.468095in}{2.078725in}}%
\pgfpathlineto{\pgfqpoint{1.509429in}{2.086776in}}%
\pgfpathlineto{\pgfqpoint{1.550762in}{2.095106in}}%
\pgfpathlineto{\pgfqpoint{1.592095in}{2.103715in}}%
\pgfpathlineto{\pgfqpoint{1.633429in}{2.112603in}}%
\pgfpathlineto{\pgfqpoint{1.674762in}{2.121769in}}%
\pgfpathlineto{\pgfqpoint{1.716095in}{2.131215in}}%
\pgfpathlineto{\pgfqpoint{1.757429in}{2.140939in}}%
\pgfpathlineto{\pgfqpoint{1.798762in}{2.150942in}}%
\pgfpathlineto{\pgfqpoint{1.871095in}{2.176344in}}%
\pgfpathlineto{\pgfqpoint{1.943429in}{2.201513in}}%
\pgfpathlineto{\pgfqpoint{2.015762in}{2.226447in}}%
\pgfpathlineto{\pgfqpoint{2.088095in}{2.251148in}}%
\pgfpathlineto{\pgfqpoint{2.160429in}{2.275614in}}%
\pgfpathlineto{\pgfqpoint{2.232762in}{2.299846in}}%
\pgfpathlineto{\pgfqpoint{2.305095in}{2.323845in}}%
\pgfpathlineto{\pgfqpoint{2.377429in}{2.347609in}}%
\pgfpathlineto{\pgfqpoint{2.449762in}{2.371139in}}%
\pgfpathlineto{\pgfqpoint{2.522095in}{2.394436in}}%
\pgfpathlineto{\pgfqpoint{2.594429in}{2.417498in}}%
\pgfpathlineto{\pgfqpoint{2.666762in}{2.440326in}}%
\pgfpathlineto{\pgfqpoint{2.739095in}{2.462920in}}%
\pgfpathlineto{\pgfqpoint{2.811429in}{2.485280in}}%
\pgfpathlineto{\pgfqpoint{2.832095in}{2.491626in}}%
\pgfpathlineto{\pgfqpoint{2.883762in}{2.508613in}}%
\pgfpathlineto{\pgfqpoint{2.935429in}{2.525354in}}%
\pgfpathlineto{\pgfqpoint{2.987095in}{2.541848in}}%
\pgfpathlineto{\pgfqpoint{3.038762in}{2.558096in}}%
\pgfpathlineto{\pgfqpoint{3.090429in}{2.574098in}}%
\pgfpathlineto{\pgfqpoint{3.142095in}{2.589853in}}%
\pgfpathlineto{\pgfqpoint{3.193762in}{2.605362in}}%
\pgfpathlineto{\pgfqpoint{3.245429in}{2.620624in}}%
\pgfpathlineto{\pgfqpoint{3.297095in}{2.635640in}}%
\pgfpathlineto{\pgfqpoint{3.348762in}{2.650409in}}%
\pgfpathlineto{\pgfqpoint{3.400429in}{2.664932in}}%
\pgfpathlineto{\pgfqpoint{3.452095in}{2.679209in}}%
\pgfpathlineto{\pgfqpoint{3.503762in}{2.693239in}}%
\pgfpathlineto{\pgfqpoint{3.555429in}{2.707023in}}%
\pgfpathlineto{\pgfqpoint{3.607095in}{2.720560in}}%
\pgfpathlineto{\pgfqpoint{3.658762in}{2.733851in}}%
\pgfpathlineto{\pgfqpoint{3.710429in}{2.746896in}}%
\pgfpathlineto{\pgfqpoint{3.762095in}{2.759694in}}%
\pgfpathlineto{\pgfqpoint{3.813762in}{2.772246in}}%
\pgfpathlineto{\pgfqpoint{3.865429in}{2.784551in}}%
\pgfpathlineto{\pgfqpoint{3.927429in}{2.798349in}}%
\pgfpathlineto{\pgfqpoint{3.989429in}{2.811873in}}%
\pgfpathlineto{\pgfqpoint{4.051429in}{2.825126in}}%
\pgfpathlineto{\pgfqpoint{4.103095in}{2.835961in}}%
\pgfpathlineto{\pgfqpoint{4.154762in}{2.846607in}}%
\pgfpathlineto{\pgfqpoint{4.206429in}{2.857063in}}%
\pgfpathlineto{\pgfqpoint{4.258095in}{2.867330in}}%
\pgfpathlineto{\pgfqpoint{4.309762in}{2.877408in}}%
\pgfpathlineto{\pgfqpoint{4.361429in}{2.887297in}}%
\pgfpathlineto{\pgfqpoint{4.413095in}{2.896996in}}%
\pgfpathlineto{\pgfqpoint{4.464762in}{2.906505in}}%
\pgfpathlineto{\pgfqpoint{4.516429in}{2.915826in}}%
\pgfpathlineto{\pgfqpoint{4.568095in}{2.924957in}}%
\pgfpathlineto{\pgfqpoint{4.619762in}{2.933898in}}%
\pgfpathlineto{\pgfqpoint{4.671429in}{2.942651in}}%
\pgfpathlineto{\pgfqpoint{4.723095in}{2.951214in}}%
\pgfpathlineto{\pgfqpoint{4.774762in}{2.959587in}}%
\pgfpathlineto{\pgfqpoint{4.826429in}{2.967771in}}%
\pgfpathlineto{\pgfqpoint{4.878095in}{2.975766in}}%
\pgfpathlineto{\pgfqpoint{4.898762in}{2.978911in}}%
\pgfpathlineto{\pgfqpoint{4.981429in}{2.989551in}}%
\pgfpathlineto{\pgfqpoint{5.064095in}{2.999991in}}%
\pgfpathlineto{\pgfqpoint{5.146762in}{3.010231in}}%
\pgfpathlineto{\pgfqpoint{5.229429in}{3.020270in}}%
\pgfpathlineto{\pgfqpoint{5.312095in}{3.030110in}}%
\pgfpathlineto{\pgfqpoint{5.394762in}{3.039749in}}%
\pgfpathlineto{\pgfqpoint{5.477429in}{3.049189in}}%
\pgfpathlineto{\pgfqpoint{5.560095in}{3.058428in}}%
\pgfpathlineto{\pgfqpoint{5.642762in}{3.067467in}}%
\pgfpathlineto{\pgfqpoint{5.725429in}{3.076307in}}%
\pgfpathlineto{\pgfqpoint{5.808095in}{3.084946in}}%
\pgfpathlineto{\pgfqpoint{5.890762in}{3.093385in}}%
\pgfpathlineto{\pgfqpoint{5.973429in}{3.101624in}}%
\pgfpathlineto{\pgfqpoint{6.056095in}{3.109662in}}%
\pgfpathlineto{\pgfqpoint{6.138762in}{3.117501in}}%
\pgfpathlineto{\pgfqpoint{6.221429in}{3.125140in}}%
\pgfpathlineto{\pgfqpoint{6.304095in}{3.132578in}}%
\pgfpathlineto{\pgfqpoint{6.386762in}{3.139817in}}%
\pgfpathlineto{\pgfqpoint{6.469429in}{3.146855in}}%
\pgfpathlineto{\pgfqpoint{6.552095in}{3.153694in}}%
\pgfpathlineto{\pgfqpoint{6.634762in}{3.160332in}}%
\pgfpathlineto{\pgfqpoint{6.717429in}{3.166770in}}%
\pgfpathlineto{\pgfqpoint{6.800095in}{3.173008in}}%
\pgfpathlineto{\pgfqpoint{6.882762in}{3.179046in}}%
\pgfpathlineto{\pgfqpoint{6.975762in}{3.185601in}}%
\pgfpathlineto{\pgfqpoint{7.058429in}{3.191226in}}%
\pgfpathlineto{\pgfqpoint{7.141095in}{3.196653in}}%
\pgfpathlineto{\pgfqpoint{7.223762in}{3.201882in}}%
\pgfpathlineto{\pgfqpoint{7.306429in}{3.206912in}}%
\pgfpathlineto{\pgfqpoint{7.389095in}{3.211743in}}%
\pgfpathlineto{\pgfqpoint{7.471762in}{3.216377in}}%
\pgfpathlineto{\pgfqpoint{7.554429in}{3.220812in}}%
\pgfpathlineto{\pgfqpoint{7.637095in}{3.225049in}}%
\pgfpathlineto{\pgfqpoint{7.719762in}{3.229087in}}%
\pgfpathlineto{\pgfqpoint{7.802429in}{3.232927in}}%
\pgfpathlineto{\pgfqpoint{7.885095in}{3.236569in}}%
\pgfpathlineto{\pgfqpoint{7.967762in}{3.240013in}}%
\pgfpathlineto{\pgfqpoint{8.050429in}{3.243258in}}%
\pgfpathlineto{\pgfqpoint{8.133095in}{3.246305in}}%
\pgfpathlineto{\pgfqpoint{8.215762in}{3.249154in}}%
\pgfpathlineto{\pgfqpoint{8.298429in}{3.251804in}}%
\pgfpathlineto{\pgfqpoint{8.381095in}{3.254256in}}%
\pgfpathlineto{\pgfqpoint{8.463762in}{3.256510in}}%
\pgfpathlineto{\pgfqpoint{8.546429in}{3.258565in}}%
\pgfpathlineto{\pgfqpoint{8.629095in}{3.260422in}}%
\pgfpathlineto{\pgfqpoint{8.711762in}{3.262081in}}%
\pgfpathlineto{\pgfqpoint{8.794429in}{3.263541in}}%
\pgfpathlineto{\pgfqpoint{8.877095in}{3.264803in}}%
\pgfpathlineto{\pgfqpoint{8.959762in}{3.265867in}}%
\pgfpathlineto{\pgfqpoint{9.052762in}{3.267033in}}%
\pgfpathlineto{\pgfqpoint{9.156095in}{3.268900in}}%
\pgfpathlineto{\pgfqpoint{9.259429in}{3.270558in}}%
\pgfpathlineto{\pgfqpoint{9.362762in}{3.272010in}}%
\pgfpathlineto{\pgfqpoint{9.466095in}{3.273253in}}%
\pgfpathlineto{\pgfqpoint{9.569429in}{3.274289in}}%
\pgfpathlineto{\pgfqpoint{9.672762in}{3.275118in}}%
\pgfpathlineto{\pgfqpoint{9.776095in}{3.275739in}}%
\pgfpathlineto{\pgfqpoint{9.879429in}{3.276152in}}%
\pgfpathlineto{\pgfqpoint{9.982762in}{3.276358in}}%
\pgfpathlineto{\pgfqpoint{10.066429in}{3.276372in}}%
\pgfpathlineto{\pgfqpoint{10.066429in}{3.276372in}}%
\pgfusepath{stroke}%
\end{pgfscope}%
\begin{pgfscope}%
\pgfpathrectangle{\pgfqpoint{0.765429in}{0.693364in}}{\pgfqpoint{9.300000in}{6.795000in}}%
\pgfusepath{clip}%
\pgfsetbuttcap%
\pgfsetroundjoin%
\pgfsetlinewidth{2.007500pt}%
\definecolor{currentstroke}{rgb}{0.858824,0.721569,0.368627}%
\pgfsetstrokecolor{currentstroke}%
\pgfsetdash{{16.000000pt}{6.000000pt}}{0.000000pt}%
\pgfpathmoveto{\pgfqpoint{0.765429in}{1.071850in}}%
\pgfpathlineto{\pgfqpoint{0.817095in}{1.072655in}}%
\pgfpathlineto{\pgfqpoint{0.868762in}{1.073689in}}%
\pgfpathlineto{\pgfqpoint{0.920429in}{1.074952in}}%
\pgfpathlineto{\pgfqpoint{0.972095in}{1.076445in}}%
\pgfpathlineto{\pgfqpoint{1.023762in}{1.078167in}}%
\pgfpathlineto{\pgfqpoint{1.075429in}{1.080118in}}%
\pgfpathlineto{\pgfqpoint{1.127095in}{1.082298in}}%
\pgfpathlineto{\pgfqpoint{1.178762in}{1.084708in}}%
\pgfpathlineto{\pgfqpoint{1.230429in}{1.087347in}}%
\pgfpathlineto{\pgfqpoint{1.282095in}{1.090216in}}%
\pgfpathlineto{\pgfqpoint{1.333762in}{1.093314in}}%
\pgfpathlineto{\pgfqpoint{1.385429in}{1.096641in}}%
\pgfpathlineto{\pgfqpoint{1.437095in}{1.100197in}}%
\pgfpathlineto{\pgfqpoint{1.488762in}{1.103983in}}%
\pgfpathlineto{\pgfqpoint{1.540429in}{1.107998in}}%
\pgfpathlineto{\pgfqpoint{1.592095in}{1.112243in}}%
\pgfpathlineto{\pgfqpoint{1.643762in}{1.116717in}}%
\pgfpathlineto{\pgfqpoint{1.695429in}{1.121420in}}%
\pgfpathlineto{\pgfqpoint{1.747095in}{1.126352in}}%
\pgfpathlineto{\pgfqpoint{1.798762in}{1.131514in}}%
\pgfpathlineto{\pgfqpoint{1.891762in}{1.144017in}}%
\pgfpathlineto{\pgfqpoint{1.984762in}{1.156743in}}%
\pgfpathlineto{\pgfqpoint{2.077762in}{1.169690in}}%
\pgfpathlineto{\pgfqpoint{2.170762in}{1.182859in}}%
\pgfpathlineto{\pgfqpoint{2.263762in}{1.196250in}}%
\pgfpathlineto{\pgfqpoint{2.356762in}{1.209863in}}%
\pgfpathlineto{\pgfqpoint{2.449762in}{1.223698in}}%
\pgfpathlineto{\pgfqpoint{2.542762in}{1.237755in}}%
\pgfpathlineto{\pgfqpoint{2.635762in}{1.252033in}}%
\pgfpathlineto{\pgfqpoint{2.728762in}{1.266534in}}%
\pgfpathlineto{\pgfqpoint{2.821762in}{1.281256in}}%
\pgfpathlineto{\pgfqpoint{2.832095in}{1.282906in}}%
\pgfpathlineto{\pgfqpoint{3.245429in}{1.354587in}}%
\pgfpathlineto{\pgfqpoint{3.658762in}{1.426047in}}%
\pgfpathlineto{\pgfqpoint{3.958429in}{1.477664in}}%
\pgfpathlineto{\pgfqpoint{4.940095in}{1.646334in}}%
\pgfpathlineto{\pgfqpoint{5.126095in}{1.678744in}}%
\pgfpathlineto{\pgfqpoint{5.301762in}{1.709134in}}%
\pgfpathlineto{\pgfqpoint{5.477429in}{1.739312in}}%
\pgfpathlineto{\pgfqpoint{5.653095in}{1.769276in}}%
\pgfpathlineto{\pgfqpoint{5.828762in}{1.799027in}}%
\pgfpathlineto{\pgfqpoint{6.004429in}{1.828565in}}%
\pgfpathlineto{\pgfqpoint{6.180095in}{1.857890in}}%
\pgfpathlineto{\pgfqpoint{6.355762in}{1.887002in}}%
\pgfpathlineto{\pgfqpoint{6.531429in}{1.915901in}}%
\pgfpathlineto{\pgfqpoint{6.707095in}{1.944587in}}%
\pgfpathlineto{\pgfqpoint{6.882762in}{1.973060in}}%
\pgfpathlineto{\pgfqpoint{6.975762in}{1.988005in}}%
\pgfpathlineto{\pgfqpoint{7.099762in}{2.007312in}}%
\pgfpathlineto{\pgfqpoint{7.223762in}{2.026390in}}%
\pgfpathlineto{\pgfqpoint{7.347762in}{2.045239in}}%
\pgfpathlineto{\pgfqpoint{7.471762in}{2.063859in}}%
\pgfpathlineto{\pgfqpoint{7.595762in}{2.082251in}}%
\pgfpathlineto{\pgfqpoint{7.719762in}{2.100413in}}%
\pgfpathlineto{\pgfqpoint{7.843762in}{2.118346in}}%
\pgfpathlineto{\pgfqpoint{7.967762in}{2.136050in}}%
\pgfpathlineto{\pgfqpoint{8.091762in}{2.153525in}}%
\pgfpathlineto{\pgfqpoint{8.215762in}{2.170771in}}%
\pgfpathlineto{\pgfqpoint{8.339762in}{2.187788in}}%
\pgfpathlineto{\pgfqpoint{8.463762in}{2.204577in}}%
\pgfpathlineto{\pgfqpoint{8.587762in}{2.221136in}}%
\pgfpathlineto{\pgfqpoint{8.711762in}{2.237466in}}%
\pgfpathlineto{\pgfqpoint{8.835762in}{2.253567in}}%
\pgfpathlineto{\pgfqpoint{8.959762in}{2.269439in}}%
\pgfpathlineto{\pgfqpoint{9.032095in}{2.278592in}}%
\pgfpathlineto{\pgfqpoint{9.125095in}{2.289119in}}%
\pgfpathlineto{\pgfqpoint{9.218095in}{2.299420in}}%
\pgfpathlineto{\pgfqpoint{9.311095in}{2.309495in}}%
\pgfpathlineto{\pgfqpoint{9.404095in}{2.319344in}}%
\pgfpathlineto{\pgfqpoint{9.497095in}{2.328967in}}%
\pgfpathlineto{\pgfqpoint{9.590095in}{2.338364in}}%
\pgfpathlineto{\pgfqpoint{9.683095in}{2.347536in}}%
\pgfpathlineto{\pgfqpoint{9.776095in}{2.356481in}}%
\pgfpathlineto{\pgfqpoint{9.869095in}{2.365201in}}%
\pgfpathlineto{\pgfqpoint{9.962095in}{2.373695in}}%
\pgfpathlineto{\pgfqpoint{10.055095in}{2.381963in}}%
\pgfpathlineto{\pgfqpoint{10.066429in}{2.382955in}}%
\pgfpathlineto{\pgfqpoint{10.066429in}{2.382955in}}%
\pgfusepath{stroke}%
\end{pgfscope}%
\begin{pgfscope}%
\pgfpathrectangle{\pgfqpoint{0.765429in}{0.693364in}}{\pgfqpoint{9.300000in}{6.795000in}}%
\pgfusepath{clip}%
\pgfsetbuttcap%
\pgfsetroundjoin%
\pgfsetlinewidth{2.007500pt}%
\definecolor{currentstroke}{rgb}{1.000000,0.000000,0.000000}%
\pgfsetstrokecolor{currentstroke}%
\pgfsetdash{{7.400000pt}{3.200000pt}}{0.000000pt}%
\pgfpathmoveto{\pgfqpoint{0.765429in}{2.645382in}}%
\pgfpathlineto{\pgfqpoint{10.065429in}{2.645382in}}%
\pgfusepath{stroke}%
\end{pgfscope}%
\begin{pgfscope}%
\pgfpathrectangle{\pgfqpoint{0.765429in}{0.693364in}}{\pgfqpoint{9.300000in}{6.795000in}}%
\pgfusepath{clip}%
\pgfsetbuttcap%
\pgfsetroundjoin%
\pgfsetlinewidth{2.007500pt}%
\definecolor{currentstroke}{rgb}{0.000000,0.000000,1.000000}%
\pgfsetstrokecolor{currentstroke}%
\pgfsetdash{{2.000000pt}{3.300000pt}}{0.000000pt}%
\pgfpathmoveto{\pgfqpoint{0.765429in}{6.184273in}}%
\pgfpathlineto{\pgfqpoint{10.065429in}{6.184273in}}%
\pgfusepath{stroke}%
\end{pgfscope}%
\begin{pgfscope}%
\pgfpathrectangle{\pgfqpoint{0.765429in}{0.693364in}}{\pgfqpoint{9.300000in}{6.795000in}}%
\pgfusepath{clip}%
\pgfsetbuttcap%
\pgfsetroundjoin%
\pgfsetlinewidth{2.007500pt}%
\definecolor{currentstroke}{rgb}{0.000000,0.501961,0.000000}%
\pgfsetstrokecolor{currentstroke}%
\pgfsetdash{{12.800000pt}{3.200000pt}{2.000000pt}{3.200000pt}}{0.000000pt}%
\pgfpathmoveto{\pgfqpoint{0.765429in}{6.870637in}}%
\pgfpathlineto{\pgfqpoint{10.065429in}{6.870637in}}%
\pgfusepath{stroke}%
\end{pgfscope}%
\begin{pgfscope}%
\pgfsetrectcap%
\pgfsetmiterjoin%
\pgfsetlinewidth{1.003750pt}%
\definecolor{currentstroke}{rgb}{0.000000,0.000000,0.000000}%
\pgfsetstrokecolor{currentstroke}%
\pgfsetdash{}{0pt}%
\pgfpathmoveto{\pgfqpoint{0.765429in}{0.693364in}}%
\pgfpathlineto{\pgfqpoint{0.765429in}{7.488364in}}%
\pgfusepath{stroke}%
\end{pgfscope}%
\begin{pgfscope}%
\pgfsetrectcap%
\pgfsetmiterjoin%
\pgfsetlinewidth{1.003750pt}%
\definecolor{currentstroke}{rgb}{0.000000,0.000000,0.000000}%
\pgfsetstrokecolor{currentstroke}%
\pgfsetdash{}{0pt}%
\pgfpathmoveto{\pgfqpoint{10.065429in}{0.693364in}}%
\pgfpathlineto{\pgfqpoint{10.065429in}{7.488364in}}%
\pgfusepath{stroke}%
\end{pgfscope}%
\begin{pgfscope}%
\pgfsetrectcap%
\pgfsetmiterjoin%
\pgfsetlinewidth{1.003750pt}%
\definecolor{currentstroke}{rgb}{0.000000,0.000000,0.000000}%
\pgfsetstrokecolor{currentstroke}%
\pgfsetdash{}{0pt}%
\pgfpathmoveto{\pgfqpoint{0.765429in}{0.693364in}}%
\pgfpathlineto{\pgfqpoint{10.065429in}{0.693364in}}%
\pgfusepath{stroke}%
\end{pgfscope}%
\begin{pgfscope}%
\pgfsetrectcap%
\pgfsetmiterjoin%
\pgfsetlinewidth{1.003750pt}%
\definecolor{currentstroke}{rgb}{0.000000,0.000000,0.000000}%
\pgfsetstrokecolor{currentstroke}%
\pgfsetdash{}{0pt}%
\pgfpathmoveto{\pgfqpoint{0.765429in}{7.488364in}}%
\pgfpathlineto{\pgfqpoint{10.065429in}{7.488364in}}%
\pgfusepath{stroke}%
\end{pgfscope}%
\begin{pgfscope}%
\definecolor{textcolor}{rgb}{1.000000,0.000000,0.000000}%
\pgfsetstrokecolor{textcolor}%
\pgfsetfillcolor{textcolor}%
\pgftext[x=1.592095in,y=2.707155in,left,base]{\color{textcolor}\rmfamily\fontsize{20.000000}{24.000000}\selectfont boson star}%
\end{pgfscope}%
\begin{pgfscope}%
\definecolor{textcolor}{rgb}{0.000000,0.000000,1.000000}%
\pgfsetstrokecolor{textcolor}%
\pgfsetfillcolor{textcolor}%
\pgftext[x=1.592095in,y=6.246046in,left,base]{\color{textcolor}\rmfamily\fontsize{20.000000}{24.000000}\selectfont fluid star}%
\end{pgfscope}%
\begin{pgfscope}%
\definecolor{textcolor}{rgb}{0.000000,0.500000,0.000000}%
\pgfsetstrokecolor{textcolor}%
\pgfsetfillcolor{textcolor}%
\pgftext[x=1.592095in,y=6.932409in,left,base]{\color{textcolor}\rmfamily\fontsize{20.000000}{24.000000}\selectfont black hole}%
\end{pgfscope}%
\begin{pgfscope}%
\definecolor{textcolor}{rgb}{0.858824,0.721569,0.368627}%
\pgfsetstrokecolor{textcolor}%
\pgfsetfillcolor{textcolor}%
\pgftext[x=9.032095in,y=3.411364in,left,base]{\color{textcolor}\rmfamily\fontsize{20.000000}{24.000000}\selectfont \(\displaystyle \bar{\alpha} = 0.05\)}%
\end{pgfscope}%
\begin{pgfscope}%
\definecolor{textcolor}{rgb}{0.149020,0.137255,0.133333}%
\pgfsetstrokecolor{textcolor}%
\pgfsetfillcolor{textcolor}%
\pgftext[x=9.032095in,y=2.917182in,left,base]{\color{textcolor}\rmfamily\fontsize{20.000000}{24.000000}\selectfont \(\displaystyle \bar{\alpha} = 0\)}%
\end{pgfscope}%
\begin{pgfscope}%
\definecolor{textcolor}{rgb}{0.858824,0.721569,0.368627}%
\pgfsetstrokecolor{textcolor}%
\pgfsetfillcolor{textcolor}%
\pgftext[x=9.032095in,y=2.052364in,left,base]{\color{textcolor}\rmfamily\fontsize{20.000000}{24.000000}\selectfont \(\displaystyle \bar{\alpha} = 1\)}%
\end{pgfscope}%
\end{pgfpicture}%
\makeatother%
\endgroup%